\documentclass[twocolumn,prx,lengthcheck,final,superscriptaddress]{revtex4}
\usepackage[latin9]{inputenc}
\usepackage{color}
\usepackage{units}
\usepackage{amsmath}
\usepackage{amssymb}
\usepackage{graphicx}
\usepackage[unicode=true,pdfusetitle,
 bookmarks=true,bookmarksnumbered=false,bookmarksopen=false,
 breaklinks=false,pdfborder={0 0 1},backref=false,colorlinks=true]
 {hyperref}

\makeatletter

\usepackage{tikz}
\usetikzlibrary{calc}

\@ifundefined{textcolor}{}
{%
 \definecolor{BLACK}{gray}{0}
 \definecolor{WHITE}{gray}{1}
 \definecolor{RED}{rgb}{1,0,0}
 \definecolor{GREEN}{rgb}{0,1,0}
 \definecolor{BLUE}{rgb}{0,0,1}
 \definecolor{CYAN}{cmyk}{1,0,0,0}
 \definecolor{MAGENTA}{cmyk}{0,1,0,0}
 \definecolor{YELLOW}{cmyk}{0,0,1,0}
}

\usepackage{graphicx}
\usepackage{dcolumn}
\usepackage{bm}
\usepackage{units}
\usepackage{textcomp}

\makeatother

\begin{document}

\title{Path integral framework for characterizing and controlling decoherence
induced by non-stationary environments on a quantum probe}

\author{Martin Kuffer}

\affiliation{Centro Atómico Bariloche, CONICET, CNEA, S. C. de Bariloche, 8400,
Argentina}

\affiliation{Instituto de Nanociencia y Nanotecnologia, CNEA, CONICET, S. C. de
Bariloche, 8400, Argentina}

\affiliation{Instituto Balseiro, CNEA, Universidad Nacional de Cuyo, S. C. de
Bariloche, 8400, Argentina}

\author{Analia Zwick}

\affiliation{Centro Atómico Bariloche, CONICET, CNEA, S. C. de Bariloche, 8400,
Argentina}

\affiliation{Instituto de Nanociencia y Nanotecnologia, CNEA, CONICET, S. C. de
Bariloche, 8400, Argentina}

\affiliation{Instituto Balseiro, CNEA, Universidad Nacional de Cuyo, S. C. de
Bariloche, 8400, Argentina}

\author{Gonzalo A. Álvarez}
\email{gonzalo.alvarez@cab.cnea.gov.ar}


\affiliation{Centro Atómico Bariloche, CONICET, CNEA, S. C. de Bariloche, 8400,
Argentina}

\affiliation{Instituto de Nanociencia y Nanotecnologia, CNEA, CONICET, S. C. de
Bariloche, 8400, Argentina}

\affiliation{Instituto Balseiro, CNEA, Universidad Nacional de Cuyo, S. C. de
Bariloche, 8400, Argentina}
\begin{abstract}
Reliable processing of quantum information is a milestone to achieve
for the deployment of quantum technologies. Uncontrolled, out-of-equilibrium
sources of decoherence need to be characterized in detail for designing
the control of quantum devices to mitigate the loss of quantum information.
However, quantum sensing of such environments is still a challenge
due to their non-stationary nature that in general can generate complex
high-order correlations. We here introduce a path integral framework
to characterize non-stationary environmental fluctuations by a quantum
probe. We found the solution for the decoherence decay of non-stationary,
generalized Gaussian processes that induce pure dephasing. This dephasing
when expressed in a suitable basis, based on the non-stationary noise
eigenmodes, is defined by the overlap of a generalized noise spectral
density and a filter function that depends on the control fields.
This result thus extends the validity to out-of-equilibrium environments,
of the similar general expression for the dephasing of open quantum
systems coupled to stationary noises. We show physical insights for
a broad subclass of non-stationary noises that are local-in-time,
in the sense that the noise correlation functions contain memory based
on constraints of the derivatives of the fluctuating noise paths.
Spectral and non-Markovian properties are discussed together with
implementations of the framework to treat paradigmatic environments
that are out-of-equilibrium, e.g. due to a quench and a pulsed noise.
We show that our results provide tools for probing the spectral and
time-correlation properties, and for mitigating decoherence effects
of out-of-equilibrium --non-stationary-- environments. 
\end{abstract}
\maketitle

\section{Introduction}

The progress on controlling quantum systems has lead to the development
of quantum technologies \citep{Kurizki2015,Acin2018,Awschalom2018,Deutsch2020}.
In order to deploy these technologies quantum information need to
be reliably processed. However, both the storage and processing of
quantum information in quantum devices suffer from decoherence, the
loss of quantum information as a function of time, that distort the
encoded information \citep{Zurek2003,Khodjasteh2009,alvarez_nmr_2010,Souza2012a,Maurer2012,Sar2012,Taminiau2014,Zwick2014,Alvarez2015,Suter2015,Wang2017,Dominguez2020}.
Uncontrolled sources of decoherence are ubiquitously present in the
environment of a quantum system, and methods for mitigating their
effects have been extensively explored \citep{Viola1999b,Viola1999a,Kofman2001,Kofman2004a,Khodjasteh2005,Uhrig2007,Du2009,Biercuk2009a,souza2011robust,Souza2012,Zwick2014,Suter2016}.
There is no universal optimal solution for protecting against decoherence,
as the detailed control to the system has to be tailored based on
the specific noise source \citep{Cywinski2008,Gordon2008,Uys2009,Alvarez2010,Clausen2010,Khodjasteh2010,Ajoy2011,Alvarez2011,Bar-Gill2012,Paz-Silva2014,Soare2014,Suter2016,Wang2017,Malinowski2017}.
A key required input for finding the optimal strategies for decoupling
the environmental effects is the detailed knowledge of the noise spectral
properties \citep{Alvarez2011,Bylander2011,Kotler2013,Sung2019,Sung2021,Wise2021}.
Most of the approaches for controlling and characterizing the decoherence
effects are developed for stationary noise fluctuations \citep{Kofman2001,Kofman2004a,Gordon2008,Clausen2010,Cywinski2008,Almog2011,Alvarez2011,Bylander2011,Benedetti2014,zwick_maximizing_2016,zwick2020precision}.
Developing methods for controlling and characterizing non-stationary
environmental fluctuations is a prerequisite to exploit the full extent
of quantum technologies at atomic- and nano-scales, where the environmental
systems are intrinsically of the many-body type and are out-of-equilibrium
\citep{Ma2014,Alvarez2015,Eisert2015,Grabert2016,Degen2017,Schweigler2017,Yang2017,Kirchberg2018,Abanin2019,Lewis-Swan2019,Landsman2019,Lukin2019,Brydges2019,Wang2019,Luepke2020,Dominguez2020,Chalermpusitarak2020,Wang2021}.

Dynamical decoupling noise spectroscopy is a promising tool for characterizing
fluctuating environments \citep{Meriles2010,Almog2011,Alvarez2011,Bylander2011,Bar-Gill2012}.
It is based on applying time-dependent control pulses to probe the
noise spectral properties. Several quantum sensing methods have been
designed to reconstruct the noise spectrum generated by semi-classical
and quantum fluctuating sources \citep{Cao2010,Alvarez2011,Paz-Silva2014,Norris2016,Frey2017,Mueller2018,Ferrie2018,Sung2019,Do2019,Luepke2020}.
However, it remains open how to interpret the extracted noise spectrum
probed by a quantum sensor when it is coupled to a complex and unknown
environment \citep{Wang2019,Chen2019,Wise2021,Martina2021}. More
importantly, there are no universal methods for determining the properties
of the natural, out-of-equilibrium environments that lead to non-stationary,
non-markovian noise fluctuation processes \citep{Polkovnikov2011,Ma2014,Grabert2016,Norris2016,Kirchberg2018,Wang2019,Chalermpusitarak2020,Wang2021}.

To deal with these outstanding problems, we here implement a path
integral framework to describe the decoherence process on a controlled
1/2-spin, quantum-probe coupled to non-stationary noise sources. We
set the path-integral framework for the most general pure dephasing
coupling with fluctuating fields that can be described by non-stationary,
generalized Gaussian processes, and find the solution for the decoherence
decay. Our results generalize the universal formula \citep{Kofman2001,Kofman2004a}
for the dephasing decay that only depends on the overlap of a noise
spectral density and a filter function that depends on the control
fields. Specifically, we demonstrate its extension to the case of
non-stationary Gaussian environments. 

This generalization allows to implement two important applications
for the deployment of quantum technologies, dynamical decoupling noise
spectroscopy for quantum sensing operations and mitigating by control
methods the decoherence effects of out-of-equilibrium environments.
Moreover, we provide simple interpretations of non-stationary noise
spectrums and time-dependent correlations for a broad subclass of
local-in-time noises and simple criteria to distinguish Markovian
from non-Markovian noise dynamics. We also show how to implement our
framework with two paradigmatic non-stationary noises, one derived
from a quenched environment where excitations suddenly start spreading
over a large number of degrees of freedom \citep{Polkovnikov2011,Alvarez2015,Eisert2015,Abanin2019,Wang2021}
and the other from a noise that acts near to a point in time \citep{Hall2012,Hall2013,Poggio2018,Zhang2021a}.
Overall we introduce a tool to characterize and control decoherence
effects of out-of-equilibrium environments providing avenues of quantum
information processing for the deployment of quantum technologies.

Our manuscript is organized as follows. In Sec. \ref{sec:Qubit-probe-interacting-with}
we describe the quantum-probe interacting with fluctuating fields.
In Sec. \ref{sec:Path-integral-framework} we introduce the path integral
framework to determine the qubit-probe dephasing induced by a general,
non-stationary Gaussian noise in the time- and frequency-domain. In
Sec. \ref{sec:Noise-spectral-density} we define a generalized noise
spectral density and control filter function in a suitable non-stationary,
noise eigenmode basis, and show that in this basis the dephasing is
given by the overlap between them. Based on this noise eigenmode basis,
we show how to implement (i) quantum sensing of the generalized noise
spectrum and (ii) optimal dynamical decoupling sequences to protect
the quantum system against decoherence. In Sec. \ref{sec:local-in-time}
we consider a broad subclass of non-stationary Gaussian noises that
are local-in-time to provide more direct physical meanings of parameters
that characterize these non-stationary noises within this path integral
framework. In Sec. \ref{sec:Non-stationary-noises} we apply the presented
tools to two paradigmatic out-of-equilibrium environments: a noise
acting near to a point of time by an analogy to a quantum harmonic
oscillator and a quench on the environment that suddenly starts an
Ornstein-Uhlenbeck diffusion process. Lastly, in Sec. \ref{sec:Summary-and-Conclusions}
are the concluding remarks. 

\section{\label{sec:Qubit-probe-interacting-with}Qubit-probe interacting
with fluctuating fields}

\begin{figure*}[t]
\begin{centering}
\includegraphics[width=2\columnwidth]{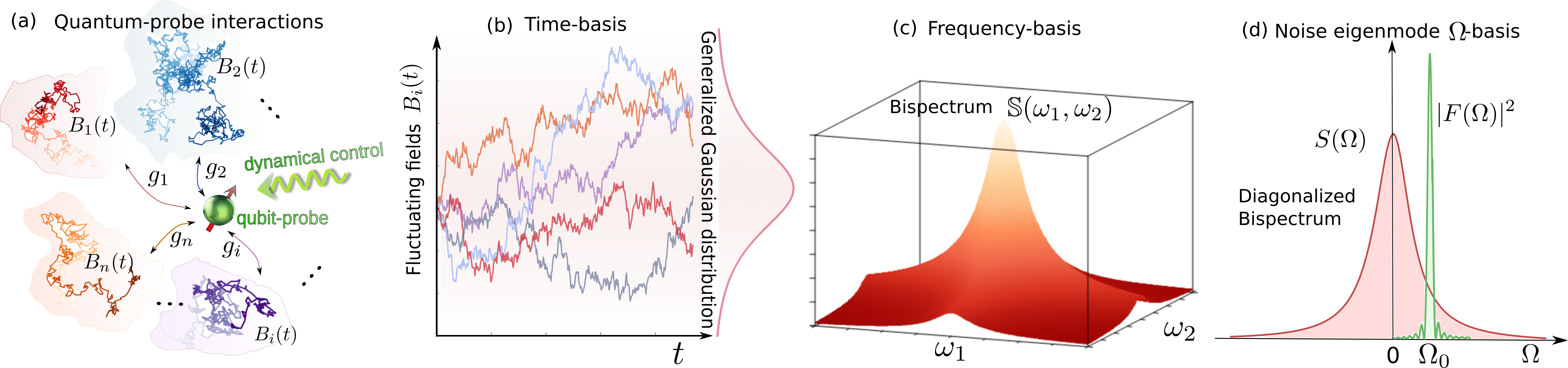}
\par\end{centering}
{\small{}\caption{\label{fig:Qprobe}Representation of the qubit-probe interaction with
fluctuating fields derived from the path integral framework. (a) Schematic
representation of a qubit-probe interacting with the fluctuating fields
$B_{i}(t)$ that generate pure dephasing. The system-bath coupling
strengths $g_{i}$ are dynamically controlled. The index $i$ labels
properties of the qubit-bath coupling network morphology. (b) Representation
of the fluctuating fields paths $B_{i}(t)$ that give a generalized
Gaussian distribution that is sensed by the quantum probe as a function
of time. The fluctuating fields are described in the time-basis $|t)$.
(c) Frequency-basis $|\omega)$ representation of the fluctuating
fields correlation functions, which provides a bidimensional, non-stationary
noise spectrum \emph{$\mathbb{S}(\omega_{1},\omega_{2})=(\omega_{1}|\mathbb{G}|\omega_{2})$.
}(d) Noise eigenmode $\Omega$-basis representation of the non-stationary,
generalized noise spectrum $S(\Omega)$ (red solid-line), obtained
after diagonalization of the noise spectrum in the frequency-basis.
A generalized filter function $F(\Omega)$ in the noise eigenmode
$\Omega$-basis is shown in green solid-line. Here, it mainly probes
a single noise eigenmode $\Omega_{0}$.}
}{\small\par}
\end{figure*}

We consider a qubit system with spin $S=1/2$ as a quantum probe experiencing
pure dephasing due to the interaction with its environment (bath)
\citep{breuer_open,Suter2016}. In the system rotating frame of reference,
the Hamiltonian is given by

\begin{equation}
\mathcal{H}=\mathcal{H}_{SB}+\mathcal{H}_{B},\quad\mathcal{H}_{SB}=S_{z}\mathbf{g}\cdot\mathbf{B},\label{eq:pure_dephasing}
\end{equation}
where $\mathcal{H}_{B}$ is the bath Hamiltonian and $\mathcal{H}_{SB}$
is a general pure dephasing system-bath interaction Hamiltonian.
The spin operator $S_{z}$ in the $z$ axis is the qubit-probe operator,
the bath-operators are represented by an $n$-dimensional vector $\mathbf{B}=\left(B_{1},\dots B_{n}\right)$,
and $\mathbf{\mathbf{g}}=\left(g_{1},\dots g_{n}\right)$ are the
system-bath coupling strengths. Notice that the index $i$ of the
components $g_{i}B_{i}$ of the system-bath interaction, labels properties
of the qubit-bath coupling network morphology. For example, it can
label different components of an hyperfine interaction tensor \citep{Ryan2010,Casanova2016},
different spins on the environment \citep{Alvarez2010,Ajoy2011,Niknam2020}
or spatial directions as in the case of anisotropic molecular diffusion
\citep{Basser1994,LeBihan2003,Alvarez2017}. Figure \ref{fig:Qprobe}(a)
shows a schematic representation of this interaction.{\small{}}
This type of interaction is encountered in a wide range of systems,
as for example nuclear spin systems in liquid and solid state NMR
\citep{Alvarez2010,Alvarez2011,Ajoy2011,Alvarez2013,AliAhmed2013},
dephasing induced by spin bearing molecules in liquid and gas state
NMR \citep{Carr1954,Meiboom1958,stepivsnik1993time,LeBihan2003,callaghan_translational_2011,Alvarez2013a},
the hyperfine interaction of electron spins in diamonds \citep{Lange2010,Ryan2010,Bar-Gill2012,Romach2015,Casanova2016},
electron spins in quantum dots \citep{Hanson2007,Cywinski2009,Barthel2010,Medford2012},
donors in silicon \citep{Kane1998}, electron spins in nanoscale nuclear
spin baths \citep{Yang2017}, superconducting qubits \citep{Bylander2011},
trapped ultra-cold atoms \citep{Sagi2010,Almog2011}, etc. 

By using an interaction representation with respect to the evolution
of the isolated environment, we eliminate the bath-Hamiltonian $\mathcal{H}_{B}$.
The system-bath Hamiltonian becomes
\begin{align}
\mathcal{H}_{SB}^{(B)}\left(t\right) & =S_{z}\mathbf{g}\cdot\left(e^{-i\mathcal{H}_{B}t}\boldsymbol{\mathbf{B}}e^{i\mathcal{H}_{B}t}\right).\label{eq:HSE-time-dependent}
\end{align}
Since $\mathcal{H}_{B}$ does not commute with $\mathcal{H}_{SB}$,
the effective system-bath interaction $\mathcal{H}_{SB}^{\left(B\right)}$
is time-dependent and the qubit-probe experiences a fluctuating coupling
to the bath. We use the semi-classical field approximation to replace
the quantum environment operator $e^{-i\mathcal{\widehat{H}}_{B}t}\mathbf{B}e^{i\mathcal{\widehat{H}}_{B}t}$
with a classical stochastic function $\mathbf{B}(t)$ {[}Fig. \ref{fig:Qprobe}
(a,b){]}. The corresponding self-correlation function is thus given
by $\langle B_{i}(t)B_{j}(t')\rangle=\frac{1}{2}\text{Tr}\left[\rho_{e}\left\{ B_{i}(t),B_{j}(t')\right\} \right]$,
where $\langle\cdot\rangle$ is the stochastic mean value of the semi-classical
stochastic fields. The rhs of the equation is the quantum correlation
function of the bath operators $B_{i/j}$, determined by the trace
$\text{Tr}$ over the bath degrees of freedom with $\rho_{e}$ as
the density matrix of the environment and $\left\{ \cdot\right\} $
denoting the anticommutator. This approximation typically called the
weak coupling approximation represents well a time-dependent quantum
correlation of the SB interaction up to second order  \citep{abragam_principles_1983,breuer_open,Kofman2001,Kofman2004a,Gordon2007,Paz-Silva2014,Norris2016}
that properly describe a wide-range of experimental setups \citep{Meriles2010,Alvarez2010,Bylander2011,Alvarez2011,Bar-Gill2012,Norris2016,Yang2017,Frey2017,Awschalom2018,Mueller2018,Sung2019,Luepke2020}.

The evolution operator of a dynamically controlled qubit-probe combined
with this fluctuating field interaction is $e^{-i\phi[\mathbf{B},\mathbf{f}]S_{z}}$,
where $\phi[\mathbf{B},\mathbf{f}]$ is the accumulated phase by the
probe during the evolution from an initial time $t_{0}$ to a final
time $t_{f}$, starting from an initial state polarized in the $XY$
plane. Specifically,
\begin{equation}
\phi[\mathbf{B},\mathbf{f}]=\int dt\sum_{i=1}^{n}f_{i}(t)B_{i}(t),\label{eq:conventional_phase}
\end{equation}
with $\int dt\equiv\int_{-\infty}^{+\infty}dt$ and the vector $\mathbf{f}(t)=(f_{1}(t),\dots,f_{n}(t))$
is a controlled modulation of each bath interaction term driven by
dynamical control techniques during $t\in[t_{0},t_{f}]$, being null
outside this time interval \citep{Viola1999a,Viola1999b,Kofman2001,Kofman2004a,Cywinski2008,Suter2016}.
For example, for a free evolution $t\in[t_{0},t_{f}]$, and $f_{i}(t)=g_{i}\Theta(t-t_{0})\Theta(t_{f}-t)$,
where $\Theta$ is the Heaviside step function, while for a dynamical
decoupling sequence of $\pi$-pulses applied to the qubit, $f_{i}(t)$
switches between $\pm g_{i}$ at the position of every pulse \citep{Cywinski2008,Suter2016}.
Notice that we write explicitly the dependence on $\mathbf{B}$ and
$\mathbf{f}$ of the qubit-probe phase, to highlight dependence of
the phase on the stochastic fluctuating field that we then describe
with the path integral approach and the control fields, respectively. 

In this article we consider the general pure dephasing system-bath
interaction Hamiltonian of Eq. (\ref{eq:pure_dephasing}). However,
the path integral framework discussed below to describe non-stationary,
general Gaussian noise that induces dephasing by this interaction,
can be extended to a dissipation interaction based on replacing the
spin-probe operator $S_{z}$ by $S_{x}$ or $S_{y}$. For this dissipation
interaction, the phase of Eq. (\ref{eq:conventional_phase}) represents
a population phase of a coherence exchange \citep{Kofman2001,Kofman2004a,Gordon2007}.
This adaptation describes interactions found for example in chemical
quantum solvation dynamics with a time-dependent structurally changing
solvent \citep{Kirchberg2018,Kirchberg2019,Kirchberg2020} and non-equilibrium
response of nano- and atomic-systems coupled to driven Caldeira-Leggett
baths \citep{Grabert2016,Reichert2016,Grabert2018}.

\section{\label{sec:Path-integral-framework}Path integral framework for the
qubit-probe dephasing}

In this section, we introduce the general path integral framework
for describing the decoherence effects on the qubit-probe, induced
by a non-stationary, general Gaussian noise that describes an out-of-equilibrium
environment. Under this framework, we define the generalized noise
spectrum for non-stationary environments in terms of the inverse of
the kernel operator that determines the probability of the noise field
paths. We then show the implementation of this framework to stationary,
general Gaussian noises, and recover the known expression for the
induced dephasing.

\subsection{Dephasing induced by non-stationary, general Gaussian noise}

We use path integrals to calculate the decoherence effects on the
qubit-probe, induced by the fluctuating field $\mathbf{B}(t)$. The
notation $\phi[\mathbf{B},\mathbf{f}]$ of the accumulated phase in
Eq. (\ref{eq:conventional_phase}), shows the functional dependence
of $\phi$ with the evolution path followed by the fluctuating field
$\mathbf{B}(t)$, where the qubit probe is driven by the control fields
$\mathbf{f}(t)$. Since the evolved phase involves real square-integrable
functions that conform a Hilbert space, we use a bra-ket notation
to simplify the calculations and interpretations. A function $f(t)$
is associated to the ket $|f)$, and similarly the multidimensional
function $\mathbf{f}(t)$ to the ket $|\mathbf{f})$. The inner product
of this space is given by 
\begin{equation}
(\mathbf{f}|\mathbf{B})\equiv\int dt\sum_{i=1}^{n}f_{i}(t)B_{i}(t).
\end{equation}
We also introduce the time-basis $\{|t):t\in\mathfrak{R}\}$ of the
Hilbert space of scalar functions of time such that $(t|f)=f(t)$
is a scalar function over the real numbers $\mathfrak{R}$. Therefore
$(t|\mathbf{f})=\left(f_{1}(t),\dots,f_{n}(t)\right)$ is a vector
function in $\mathfrak{R}^{n}$. This is analogous to the position
basis $\{|\mathbf{x}):\mathbf{x}\in\mathfrak{R}^{n}\}$ in quantum
mechanics. Then the qubit-probe phase of Eq. (\ref{eq:conventional_phase})
becomes 
\begin{equation}
\phi[\mathbf{B},\mathbf{f}]=(\mathbf{f}|\mathbf{B}),
\end{equation}
considering that \textbf{$\mathbf{f}(t)=(t|\mathbf{f})$} and \textbf{$\mathbf{B}(t)=(t|\mathbf{B})$}.

The qubit dephasing-decay is given by the ensemble average $\langle e^{i\phi[\mathbf{B},\mathbf{f}]}\rangle$
taken over all possible field path fluctuations of $\mathbf{B}$.
Figure \ref{fig:Qprobe}(b) shows a schematic representation of the
paths of the field fluctuations. Based on path integral theory, this
ensemble average is
\begin{equation}
\langle e^{i\phi[\mathbf{B},\mathbf{f}]}\rangle=\mathcal{\int D}\mathbf{B}\exp\left\{ -\mathcal{A}[\mathbf{B}]\right\} \exp\left\{ -i(\mathbf{f}|\mathbf{B})\right\} ,
\end{equation}
where $\int\mathcal{D}\mathbf{B}$ is a functional integral with the
probability distribution of $\mathbf{B}$ given by $\exp\left\{ -\mathcal{A}[\mathbf{B}]\right\} $,
and $\mathcal{A}[\mathbf{B}]$ is the action, in analogy with quantum
path integrals in imaginary time \citep{zinn}.

We consider the most general non-stationary Gaussian noise process,
for which the action is 
\begin{equation}
\mathcal{A}[\mathbf{B}]=\frac{1}{2}(\mathbf{B}|\mathbb{D}|\mathbf{B})=\frac{1}{2}\int dt\,\int dt'\,\mathbf{B}^{\dagger}(t)\mathbb{D}(t,t')\mathbf{B}(t'),\label{eq:def_gral_accion}
\end{equation}
where $\dagger$ denotes hermitian conjugation and $\mathbb{D}(t,t')=(t|\mathbb{D}|t')$
is a $n\times n$ matrix-valued function of $t$ and $t'$ that gives
the kernel of the operator $\mathbb{D}$ in terms of times. The kernel
operator $\mathbb{D}$ must be \emph{positive definite }and \emph{real}
for the path integral to be well defined, and for the action to define
a probability density $\exp\left\{ -\mathcal{A}[\mathbf{B}]\right\} $.
We assume that $\mathbb{D}$ is \emph{hermitian} without loss of generality,
such that $\mathbb{D}(t,t')=\mathbb{D}^{\dagger}(t',t)$ (see Appendix
\ref{sec:Path-integral-for}).

The dephasing-decay of the qubit-probe can be solved exactly using
path integrals for this general non-stationary Gaussian noise process
as \citep{zinn}
\begin{multline}
\langle e^{i\phi[\mathbf{B},\mathbf{f}]}\rangle=\exp\left\{ -\frac{1}{2}(\mathbf{f}|\mathbb{G}|\mathbf{f})\right\} \\
=\exp\left\{ -\frac{1}{2}\int dt\,\int dt'\,\mathbf{f}^{\dagger}(t)\mathbb{G}(t,t')\mathbf{f}(t')\right\} ,\label{eq:dephasing-decay_f-G}
\end{multline}
where the operator $\mathbb{G}$ defines the self-correlation functions
of the stochastic process
\begin{equation}
(t|\mathbb{G}|t^{\prime})_{ij}=\mathbb{G}_{ij}(t,t')=\langle B_{i}(t)B_{j}(t')\rangle.\label{eq:correlation_functions}
\end{equation}
The operator $\mathbb{G}$ is defined by the inverse of the kernel
operator 
\begin{equation}
\mathbb{G}=\mathbb{D}^{-1}\,,\label{eq:Def_G_operatorial}
\end{equation}
or equivalently by the expression
\begin{equation}
(t_{1}|\mathbb{D\,G}|t_{2})=\int dt\,\mathbb{D}(t_{1},t)\mathbb{G}(t,t_{2})=\mathbb{I}\delta(t_{1}-t_{2})\,,\label{eq:Def_Kernel_G}
\end{equation}
where $\delta(t)$ is the Dirac delta and $\mathbb{I}$ is an identity
matrix of dimension $n\times n$.

Using the bra-ket notation, one can also solve Eq. (\ref{eq:dephasing-decay_f-G})
in the frequency-basis of scalar functions $\{|\omega)=\int\frac{dt}{\sqrt{2\pi}}e^{-i\omega t}|t):\omega\in\mathfrak{R}\}$,
rather than performing the calculation using the time-basis $\{|t):t\in\mathfrak{R}\}$.
This frequency-basis, yields to the Fourier representation of the
ensemble average of the qubit-probe signal
\begin{multline}
\langle e^{i\phi[\mathbf{B},\mathbf{f}]}\rangle=\exp\left\{ -\frac{1}{2}(\mathbf{f}|\mathbb{G}|\mathbf{f})\right\} \\
=\exp\left\{ -\frac{1}{2}\int d\omega_{1}\,\int d\omega_{2}\,\mathbf{F}^{\dagger}(\omega_{1})\mathbb{S}(\omega_{1},\omega_{2})\mathbf{F}(\omega_{2})\right\} .\label{eq:dephasing-decay_f-G-freq}
\end{multline}
Here \emph{$\mathbb{S}(\omega_{1},\omega_{2})=(\omega_{1}|\mathbb{G}|\omega_{2})$
}defines a spectral density of two frequency dimensions equivalent
to the double Fourier transform of the correlation functions $\mathbb{G}(t,t')$
{[}Fig. \ref{fig:Qprobe}(c){]}, and $\mathbf{F}(\omega)=(\omega|\mathbf{f})=\left(F_{1}(\omega),\dots,F_{n}(\omega)\right)$
is a filter function defined by the Fourier transform of 
\begin{equation}
\mathbf{f}(t)=\int\frac{d\omega}{\sqrt{2\pi}}\mathbf{F}(\omega)e^{i\omega t}.
\end{equation}

Equations (\ref{eq:dephasing-decay_f-G}) and (\ref{eq:dephasing-decay_f-G-freq})
are the qubit dephasing solutions for general --non-stationary--
gaussian noises derived from the presented path integral framework.
Using the frequency-basis, Eq. (\ref{eq:Def_G_operatorial}) becomes
\begin{equation}
(\omega_{1}|\mathbb{D\,G}|\omega_{2})=\int d\omega\,\mathbb{D}(\omega_{1},\omega)\mathbb{S}(\omega,\omega_{2})=\mathbb{I}\delta(\omega_{1}-\omega_{2}),\label{eq:identity_relation_non_stationary}
\end{equation}
where $\mathbb{D}(\omega,\omega')=(\omega|\mathbb{D}|\omega')$. Therefore,
this shows that the noise generated by the non-stationary environment
is completely determined by the bispectrum $\mathbb{S}(\omega_{1},\omega_{2})$
that is defined by the inverse of the Kernel operator $\mathbb{D}$
that describes the probability density of the field paths. Similar
bispectra, functions of two frequencies, were shown to be a useful
tool to characterize \emph{stationary, non-Gaussian} noises \citep{Norris2016,Sung2019}.
Here, we show that they are also useful tools to characterize \emph{non-stationary,
Gaussian} noises.

\subsection{Dephasing induced by stationary, general Gaussian noise}

For stationary noise, the kernel $\mathbb{D}(t,t')$ and the correlation
functions $\mathbb{G}(t,t')$ are invariant under a time translation,
therefore they only depend on the time difference $t-t'$. We can
consider then that $\mathbb{D}(t,t')=\mathbb{D}(t-t')$ and $\mathbb{G}(t,t')=\mathbb{G}(t-t')$,
and the two dimensional Fourier transform of $\mathbb{G}(t,t')$ is
then block diagonal in the frequency-basis $(\omega|\mathbb{G}|\omega')=\mathbb{S}(\omega)\delta(\omega-\omega')$.
Similarly, the Fourier transform of $\mathbb{D}(t,t')$ is $(\omega|\mathbb{D}|\omega')=\mathbb{D}(\omega)\delta(\omega-\omega')$.
Equation (\ref{eq:dephasing-decay_f-G-freq}) then becomes

\begin{equation}
\langle e^{i\phi[\mathbf{B},\mathbf{f}]}\rangle=\exp\left[-\frac{1}{2}\int d\omega\,\mathbf{F}^{\dagger}(\omega)\mathbb{S}(\omega)\mathbf{F}(\omega)\right].\label{eq:gral-dec-w}
\end{equation}

According to Eq. (\ref{eq:Def_G_operatorial}), the inverse of the
Kernel operator that describes the probability density of Eq. (\ref{eq:def_gral_accion}),
defines the noise spectrum for stationary noise
\begin{equation}
\mathbb{S}(\omega)=\mathbb{D}(\omega)^{-1}\label{eq:spectral_density_stationary}
\end{equation}
when expressed in the frequency-basis. Thus, the noise spectrum of
stationary noises is further simplified with respect to Eq. (\ref{eq:identity_relation_non_stationary})
to the multiplicative inverse of the operator $\mathbb{D}(\omega)$
in this basis, i.e. it is the inverse for each frequency mode.

Equation (\ref{eq:gral-dec-w}) expresses the dephasing-decay induced
by a stationary noise in terms of an integral over only one frequency,
in contrast with Eq. (\ref{eq:dephasing-decay_f-G-freq}) that integrates
an overlap between the filter function and the spectral density of
a non-stationary noise over two different frequencies. As a result,
we here recovered the universal formula for the qubit dephasing under
the weak coupling approximation mainly considered for the $n=1$ case,
i.e. when the dephasing is given by $\langle e^{i\phi[\mathbf{B},\mathbf{f}]}\rangle=\exp\left[-\frac{1}{2}\int d\omega\,S(\omega)\left|F(\omega)\right|^{2}\right]$
\citep{Kofman2001,Kofman2004a,Gordon2008,Biercuk2009,Uys2009,Clausen2010,Pasini2010,Ajoy2011}.
This formula for the dephasing is typically used for designing optimal
control \citep{Kofman2004a,Cywinski2008,Ajoy2011} and for dynamical
decoupling noise spectroscopy \citep{Alvarez2011,Bylander2011,Meriles2010,Almog2011,Loretz2013,Yan2013,Bar-Gill2012,Kotler2013}.

\section{\label{sec:Noise-spectral-density}Noise spectral density on the
noise eigenmode basis}

In this section, we first introduce the noise eigenmode basis that
defines the proper basis for non-stationary noises, to generalize
the universal formula for the dephasing of open quantum systems that
depends on the overlap between a noise spectral density and a qubit-control
filter function. We then show how stationary noises are described
in this noise eigenmode basis. Based on the eigenmode representation,
we discuss two important applications of this dephasing generalization
to non-stationary environments: we show how to implement dynamical
decoupling noise spectroscopy for quantum sensing, and optimized control
methods to mitigate decoherence effects.

\subsection{Non-stationary noise modes\label{subsec:Non-stationary-noise-modes}}

The kernel operator $\mathbb{D}$ is in general \emph{non-stationary},
and therefore the frequency-basis does not \emph{diagonalize it},
leading to the bispectrum $\mathbb{S}(\omega_{1},\omega_{2})$ of
Eq. (\ref{eq:dephasing-decay_f-G-freq}). For this reason, in general
the frequency-basis $\left\{ |\omega):\omega\in\mathfrak{R}\right\} $
are not the \emph{eigenfunctions} of $\mathbb{D}$ for non-stationary
noises. As the kernel operator $\mathbb{D}$ is hermitian, we can
introduce the basis $\left\{ |\Omega):\Omega\in I\right\} $ that
diagonalizes it
\begin{equation}
\mathbb{D}|\Omega)=D(\Omega)|\Omega),\label{eq:def_modos_no_estacionarios}
\end{equation}
where $I$ is a set that indexes the elements of the eigenmode-basis
and $D(\Omega)\in\mathfrak{R}_{>0}$ are the corresponding eigenvalues.
This basis is normalized to satisfy the completeness relation $\mathbb{I}=\int d\Omega\,|\Omega)(\Omega|$,
where $\int d\Omega=\int_{I}d\Omega$ represents an integral when
the spectrum is continuous or a sum when it is discrete. Notice that
$|\Omega)$ are \emph{vectors }in the same space as $|\mathbf{B})$,
and they represent the non-stationary modes of the noise fluctuations
$(t|\mathbf{B})=\mathbf{B}(t)=\int d\Omega\,b_{\Omega}(t|\Omega)$,
where $(t|\Omega)$ is a vector-valued time dependent function and
$b_{\Omega}=(\Omega|\mathbf{B})$ are scalars.

The correlation operator $\mathbb{G}$ is then diagonal in this basis
$\mathbb{G}|\Omega)=\left[D(\Omega)\right]^{-1}|\Omega)$, as $\mathbb{I}|\Omega)=\mathbb{GD}|\Omega)=\mathbb{G}|\Omega)D(\Omega)$
following Eq. (\ref{eq:Def_G_operatorial}) and considering that $\left\{ |\Omega):\Omega\in I\right\} $
is a basis. As a result, we can now define a \emph{generalized noise
spectrum} for \emph{non-stationary} Gaussian noises as the eigenvalues
of the correlation operator
\begin{equation}
S(\Omega)=\left[D(\Omega)\right]^{-1},\label{eq:spectral_density_lamda}
\end{equation}
where $\mathbb{G}|\Omega)=S(\Omega)|\Omega)$. The attenuation argument
of the dephasing-decay of Eq. (\ref{eq:dephasing-decay_f-G}) is then
\begin{equation}
(\boldsymbol{f}|\mathbb{G}|\boldsymbol{f})=\int d\Omega\,S(\Omega)|(\Omega|\boldsymbol{f})|^{2}
\end{equation}
 in this noise eigenmode basis. This $\Omega$\emph{-}basis provides
then a natural way to define a \emph{generalized filter function}
\emph{for these non-stationary noises} as $F(\Omega)=(\Omega|\boldsymbol{f})$,
to obtain 
\begin{equation}
\langle e^{i\phi[\mathbf{B},\mathbf{f}]}\rangle=\exp\left[-\frac{1}{2}\int d\Omega\,S(\Omega)\left|F(\Omega)\right|^{2}\right].\label{eq:Dec_lambda}
\end{equation}

The \emph{generalized noise spectrum} $S(\Omega)$ for a \emph{non-stationary
Gaussian noise} is thus the \emph{multiplicative inverse }of the \emph{eigenvalues
of the kernel operator} $\mathbb{D}$, indexed by the noise eigenmode
parameter $\Omega$. 

Notice that as $\Omega$ is a parameter, in general its physical meaning
depends specifically on the non-stationary process, and on the parametrization
of the functional behaviour $S(\Omega)$, as it is always possible
to reparametrize the eigenmodes by a different parameter. Conversely,
the eigenvalues of the operator $\mathbb{D}$ and their corresponding
noise eigenspaces are always independent of the parametrization. By
identifying the basis that diagonalizes the kernel operator $\mathbb{D}$,
we have mapped the bidimensional spectral density that depends on
two frequencies {[}Fig. \ref{fig:Qprobe}(c){]} into a mono-parametrical
spectral density based on the non-stationary noise eigenmodes {[}Fig.
\ref{fig:Qprobe}(d){]}. This sets one of the main results of this
article, where Eq. (\ref{eq:Dec_lambda}) is a generalization \emph{for
the qubit-probe dephasing }for \emph{non-stationary Gaussian noise}
that is only determined by the overlap between the generalized spectral
density of the environment $S(\Omega)$ and the generalized filter
function $\left|F(\Omega)\right|^{2}$ determined by the control on
the qubit-probe. This result thus extends the validity to non-stationary
noises, of the universal formula for stationary pure dephasing noises
of open quantum systems within the weak coupling approximation \citep{Kofman2001,Kofman2004a,Gordon2008,Biercuk2009,Uys2009,Clausen2010,Pasini2010,Ajoy2011,Alvarez2011}.

\subsection{Stationary noise modes\label{subsec:Stationary-noise-modes}}

In the stationary noise case, according to Eq. (\ref{eq:gral-dec-w}),
the frequency dimension is separable from the morphological dimension
of the qubit-bath coupling network represented by the vector indices
$i$ of $\mathbf{B}$ and $\mathbf{F}$ in Eq. (\ref{eq:correlation_functions}),
where we obtain $\mathbb{D}|\omega)=\mathbb{D}(\omega)|\omega)$.
The noise eigenmodes in the stationary case are defined by the eigenvectors
$\mathbf{b}_{m}(\omega)\in\mathfrak{R}^{n}$ of $\mathbb{D}(\omega)$,
with $m=1,\dots,n$, such that $\mathbb{D}(\omega)\mathbf{b}_{m}(\omega)=D_{m}(\omega)\mathbf{b}_{m}(\omega)$
with $D_{m}(\omega)\in\mathfrak{R}_{>0}$. Therefore the eigenmodes
of the Kernel operator are $|\omega,m)=\mathbf{b}_{m}(\omega)|\omega)$,
and it acts on them as $\mathbb{D}|\omega,m)=D_{m}(\omega)|\omega,m)$.
We consider the eigenmode basis index as $\Omega=(\omega,m)$ and
use the notation $D(\Omega)=D_{m}(\omega)$. The dephasing of Eq.
(\ref{eq:Dec_lambda}) then becomes

\begin{equation}
\langle e^{i\phi[\mathbf{B},\mathbf{f}]}\rangle=\exp\left[-\frac{1}{2}\sum_{m}\int d\omega\,S_{m}(\omega)\left|F_{m}(\omega)\right|^{2}\right],\label{eq:decoherence_overlap_frequency_modes_stationary}
\end{equation}
where $S_{m}(\omega)=D_{m}^{-1}(\omega)$ are the eigenspectrums of
the bath fluctuations on the qubit-probe. The index $m$ for example
defines the principal axes of anisotropic diffusion tensors in magnetic
resonance imaging \citep{Basser1994,LeBihan2003,Alvarez2017}.

While the choice of the frequency $\omega$ as the relevant parameter
to describe the noise eigenmodes for the stationary case seems natural,
its selection is also arbitrary as the choice of $\Omega$ in general.
For example one can keep the diagonal form of Eq. (\ref{eq:decoherence_overlap_frequency_modes_stationary})
but use, instead of the angular frequency $\omega$, the frequency
$\nu=\omega/(2\pi)$. One can also use a different reparamerization
like $\lambda=\omega^{\nicefrac{1}{3}}$, or replace the complex exponentials
with trigonometric functions and describe the noise eigenmodes with
two parameters $(E,\pi)$, with $E=\omega^{2}$ and $\pi$ the parity
of the trigonometric funtions (see Appendix \ref{sec:Noise-eigenmode-basis}).

\subsection{Dynamical decoupling noise spectroscopy of non-stationary environments\label{subsec:Dynamical-decoupling-noise}}

An important application of the presented framework, that leads to
the generalized picture for the qubit-probe dephasing based on Eqs.
(\ref{eq:dephasing-decay_f-G-freq}) and (\ref{eq:Dec_lambda}), is
that they can be used to probe the noise spectral properties for non-stationary,
general Gaussian noises. Several methods have been designed to probe
the noise spectrum of stationary noises based on Eq. (\ref{eq:gral-dec-w}),
mainly for the $n=1$ case, i.e. when the dephasing is given by $\langle e^{i\phi[\mathbf{B},\mathbf{f}]}\rangle=\exp\left[-\frac{1}{2}\int d\omega\,S(\omega)\left|F(\omega)\right|^{2}\right]$
\citep{Alvarez2011,Bylander2011,Meriles2010,Almog2011,Loretz2013,Yan2013,Bar-Gill2012,Kotler2013}.
Control sequences generate filter functions that can allow only specific
frequency components of the spectral density to produce dephasing
on the qubit-probe system. The width of these ``pass bands'' filters
can be made arbitrarily narrow. Therefore the spectral density can
be reconstructed by performing series of measurement with different
filter functions to scan the noise spectrum. This procedure termed
dynamical decoupling noise spectroscopy \citep{Alvarez2011} can be
performed either by using continuous fields \citep{Almog2011,Loretz2013,Yan2013}
or sequences of pulses \citep{Meriles2010,Alvarez2011,Bylander2011}.

The noise eigenmodes $\left|\Omega\right)$ that diagonalize the Kernel
operator $\mathbb{D}$ are thus the natural basis to probe the non-stationary
noise spectrum based on Eq. (\ref{eq:Dec_lambda}). In the simplest
case of stationary noise, this basis becomes $\left|\omega,m\right)$,
and therefore using a modulating function $f_{m}(t)=g_{m}e^{-i\omega_{0}t}\propto\left(t|\omega_{0}\right)$
driven by control fields, one can probe the frequency modes of the
noise spectrum for each morphological eigenmode $m$ by scanning $\omega_{0}$.
For such control modulations, the filter function $|F_{m}(\omega)|^{2}\propto\delta(\omega-\omega_{0})$
senses single frequency modes according to Eq. (\ref{eq:decoherence_overlap_frequency_modes_stationary}),
and leads to the so-called continuous-wave (CW) noise spectroscopy
\citep{Slichter1964,Ailion1965,Look1966,Almog2011,Loretz2013,Yan2013}.
However, for non-stationary noises the frequency-basis in general
cannot probe selectively the noise eigenmodes. In this case, the eigenvectors
$\left|\Omega\right)$ define the proper basis to probe the single
noise eigenmodes according to Eq. \textbf{(}\ref{eq:def_modos_no_estacionarios}\textbf{)}.
Designing filter functions $F(\Omega)$ such that $|F(\Omega)|^{2}\propto\delta(\Omega-\Omega_{0})$
{[}Fig. \ref{fig:Qprobe}(d){]} based on finding the control modulation
functions that satisfy $\mathbf{f}_{\Omega_{0}}(t)=\left(t|\Omega_{0}\right)$,
one can probe the noise eigenspectrum by scanning it by changing $\Omega_{0}$
based on Eq. (\ref{eq:Dec_lambda}) as $(\boldsymbol{f}|\mathbb{G}|\boldsymbol{f})\propto S(\Omega_{0})$.
Therefore, if the $\Omega$-basis is known, dynamical decoupling noise
spectroscopy approaches designed to scan the spectral density in the
frequency domain for stationary noises \citep{Alvarez2011,Bylander2011,Almog2011},
can now be adapted to probe non-stationary noises using the $\Omega$-domain.

Alternatively, if information about the $\Omega$-basis is not known,
dynamical decoupling noise spectroscopy to scan multifrequency spectral
densities can be implemented based on the general form derived in
Eq. (\ref{eq:dephasing-decay_f-G-freq}). Again, qubit noise spectroscopy
control methods for estimating high-order noise spectra (so-called
polyspectra), have been developed for stationary non-gaussian noises
\citep{Norris2016,Sung2019}. The method is based on using dynamical
control approaches based on frequency comb control modulations to
design multidimensional filter functions to probe the polyspectra
via repetition of suitable pulse sequences. Based on Eq. (\ref{eq:dephasing-decay_f-G-freq}),
this technique can now be straightforwardly adapted to estimate the
non-stationary bispectrum $\mathbb{S}(\omega_{1},\omega_{2})$ that
induces the qubit-probe dephasing. Then, once the bispectrum is estimated,
it can then be diagonalized to determine its eigenmode basis $\left|\Omega\right)$
and eigenspectrum $S(\Omega)$. 

\subsection{Mitigating decoherence effects of non-stationary noises}

The reduction of decoherence effects on a qubit-probe coupled to
non-stationary --out of equilibrium-- noises is other important
application of the presented framework. The key result for attaining
this goal, is the extension of the universal expression for the qubit
dephasing induced by non-stationary, general Gaussian noises based
on the overlap between a noise spectral density and a control filter
function as demonstrated with Eq. (\ref{eq:Dec_lambda}). This extension
allows implementing optimal control methods that were developed for
mitigating decoherence induced by stationary noises on open quantum
systems. These methods are based on finding an optimal control filter
function $\mathbf{F}_{\text{opt}}(\omega)=(\omega|\mathbf{f}_{\text{opt}})$
that minimizes the overlap between the noise spectral density and
the control filter function in Eq. (\ref{eq:gral-dec-w}), and then
implementing it experimentally in the time-basis as $\mathbf{f}_{\text{opt}}(t)=(t|\mathbf{\mathbf{f}}_{\text{opt}})$
\citep{Gordon2008,Uys2009,Clausen2010,Zwick2014}.

These control strategies can now be applied to non-stationary noises
by designing an optimal filter $F_{\text{opt}}(\Omega)$ that, analogously
to the stationary case, minimizes the overlap with the non-stationary
Gaussian noise spectrum $S(\Omega)$ in Eq. (\ref{eq:Dec_lambda}).
This is only possible due to the introduction of the generalized non-stationary
Gaussian noise spectrum based on determining the noise eigenmodes
$\mathbb{G}|\Omega)=S(\Omega)|\Omega)$, and is one of the main results
of this paper. As described in Sec. \ref{subsec:Dynamical-decoupling-noise},
$S(\Omega)$ can be inferred, so as to determine the proper modulation
control $|\mathbf{f}_{\text{opt}})$ that provides the optimal filter
$(\Omega|\mathbf{f}_{\text{opt}})=F_{\text{opt}}(\Omega)$. This
control, can then be expressed in the time-basis as $\mathbf{f}_{\text{opt}}(t)=(t|\mathbf{f}_{\text{opt}})$
for its experimental implementation, as one does to decouple stationary
environments. Therefore we have shown here, how the presented generalized
path integral framework can be used for implementing dynamical control
methods to mitigate decoherence effects induced by non-stationary,
general Gaussian noises. 

\section{Non-stationary noise processes local-in-time\label{sec:local-in-time}}

In this section we consider a broad subclass of non-stationary Gaussian
noises that are local-in-time, as they allow more direct interpretations
of the physical meaning that provides the path integral approach for
the noise processes. We show that the dephasing for this type of non-stationary
noises can be described by a differential operator based on constraints
to the derivatives of the fluctuating field paths. We also show how
these constraints are reflected on the functional behavior of the
non-stationary noise spectrums. Therefore, the qubit-probe dephasing
can be obtained by solving ordinary differential equations rather
than integral equations as is the case for non local-in-time noises.
This picture allows to obtain simpler expression for the noise spectrum
of local-in-time, stationary noises, as the inverse of the differential
operator. In this case the differential operator is determined by
a matrix polynomial of the frequency modes. We also show how this
description for non-stationary, local-in-time noises, provides conditions
for differentiating Markovian noises from non-Markovian ones. Moreover,
we introduce a generalized Markovian process which includes all the
derivatives of $\mathbf{B}$ in the stochastic process, that fully
describe local-in-time non-stationary processes. The state of $\mathbf{B}$
is not only determined by the \emph{probability distribution} of $\mathbf{B}$,
but by the \emph{joint probability distribution} of $\mathbf{B}$
and its derivatives. These local-in-time noises while in general can
be non-Markovian, in our generalized framework appear as a natural
extension of Markovian noises, and maintain many of their properties
while being able to model a greater variety of environments.

\subsection{Local-in-time framework\label{subsec:Local-in-time-framework}}

We here consider a broad subclass of non-stationary Gaussian noises
that are local-in-time. The kernel that defines the probability distribution
for the possible paths that can take a local-in-time fluctuation $\mathbf{B}(t)$,
satisfies $\mathbb{D}(t,t')=0$ when $t\neq t'$. The most general
Kernel operator satisfying this condition can then be expressed as
an expansion series
\begin{equation}
\mathbb{D}(t,t')=\sum_{k=0}^{\infty}\mathbb{C}_{k}(t)\delta^{(k)}(t-t'),\label{eq:local-in-time_kernel}
\end{equation}
where $\delta^{(k)}$ denotes the $k$-th time derivative of the Dirac
delta function and $\mathbb{C}_{k}(t)$ are time dependent matrices.
When the sum is finite, the most general action satisfying these conditions
can be written as\emph{
\begin{equation}
\mathcal{A}[\mathbf{B}]=\int dt\,\sum_{k=0}^{N}\sum_{l=0}^{k}\mathbf{B}^{\dagger(k)}(t)\mathbb{D}_{k,l}(t)\mathbf{B}^{(l)}(t),\label{eq:def_operador_diferencial_no_simplificado}
\end{equation}
}where $\mathbf{B}^{(k)}(t)$ is the $k$-th time derivative of $\mathbf{B}(t)$,
$N$ is the highest derivative order that contributes to the action,
and $\mathbb{D}_{k,l}(t)$ are $n\times n$ matrices. We call \emph{local-in-time}
this kind of non-stationary noise processes. The fact that these local-in-time
noises are described by a Kernel operator $\mathbb{D}(t,t')$ that
vanishes for $t\neq t'$, indicates that long-time correlations of
the process do not exist, and therefore the correlation functions
$\mathbb{G}_{ij}(t,t')=\langle B_{i}(t)B_{j}(t')\rangle$ decay with
$\left|t'-t\right|$.

We demonstrate in Appendix \ref{sec:Simplification-Expression-D}
that any local-in-time process can be described by an action of the
form 
\begin{multline}
\mathcal{A}[\boldsymbol{\mathbf{B}}]=\frac{1}{2}(\mathbf{B}|\mathbb{D}|\mathbf{B})=\int dt\,\sum_{k=0}^{N}\mathbf{B}^{\dagger(k)}(t)\mathbb{D}_{k}^{H}(t)\mathbf{B}^{(k)}(t)\,\\
+\int dt\sum_{k=1}^{N}\mathbf{B}^{\dagger(k)}(t)\mathbb{D}_{k}^{A}(t)\mathbf{B}^{(k-1)}(t),\label{eq:local-in-time-action}
\end{multline}
where $\mathbb{D}_{k}^{H}(t)$ are $n\times n$ real symmetric (Hermitian)
matrices and $\mathbb{D}_{k}^{A}(t)$ are real antisymmetric (anti-Hermitian)
matrices. These matrices define constraints on the values that the
fluctuating field $\mathbf{B}$ and its derivatives can take. For
example, the larger $|\mathbb{D}_{k}^{H}|$ is, the smaller the derivative
$\mathbf{B}^{(k)}$ must be or equivalently, the slower $\mathbf{B}^{(k-1)}$
may change. The matrix norm $|\mathbb{D}_{k}^{H}|$ thus limits how
fast $\mathbf{B}^{(k-1)}$ can lose information of its previous state.
In particular, for $N>1$, information can now be stored in the derivatives
of $\mathbf{B}$, and it is this information storage that leads to
these processes being \emph{non-Markovian.} Based on Eq. (\ref{eq:local-in-time_kernel}),
to achieve to Eq. (\ref{eq:local-in-time-action}) we have introduced
the differential operator
\begin{multline}
\mathbb{D}(t)=\sum_{k=0}^{N}\overset{\leftarrow}{\partial_{t}^{\,k}}\mathbb{D}_{k}^{H}(t)\,\partial_{t}^{\,k}\\
+\frac{1}{2}\sum_{k=1}^{N}\left[\overset{\leftarrow}{\partial_{t}^{\,k}}\mathbb{D}_{k}^{A}(t)\partial_{t}^{\,k-1}-\overset{\leftarrow}{\partial_{t}^{\,k-1}}\mathbb{D}_{k}^{A}(t)\partial_{t}^{\,k}\right],\label{eq:local_in_time_kernel_operator}
\end{multline}
that gives the Kernel operator for local-in-time noise processes,
where $\overset{\leftarrow}{\partial_{t}}$ denotes left-wise differentiation,
such that $f(t)\overset{\leftarrow}{\partial_{t}}g(t)=f'(t)g(t)$.
Notice that if $\mathbf{B}$ is a one dimensional process, $\mathbb{D}_{k}^{A}=0$
for all $k$, since there are no non-zero real anti-Hermitian $1\times1$
matrices (scalars).

\begin{figure}
\begin{centering}
\includegraphics[width=1\columnwidth]{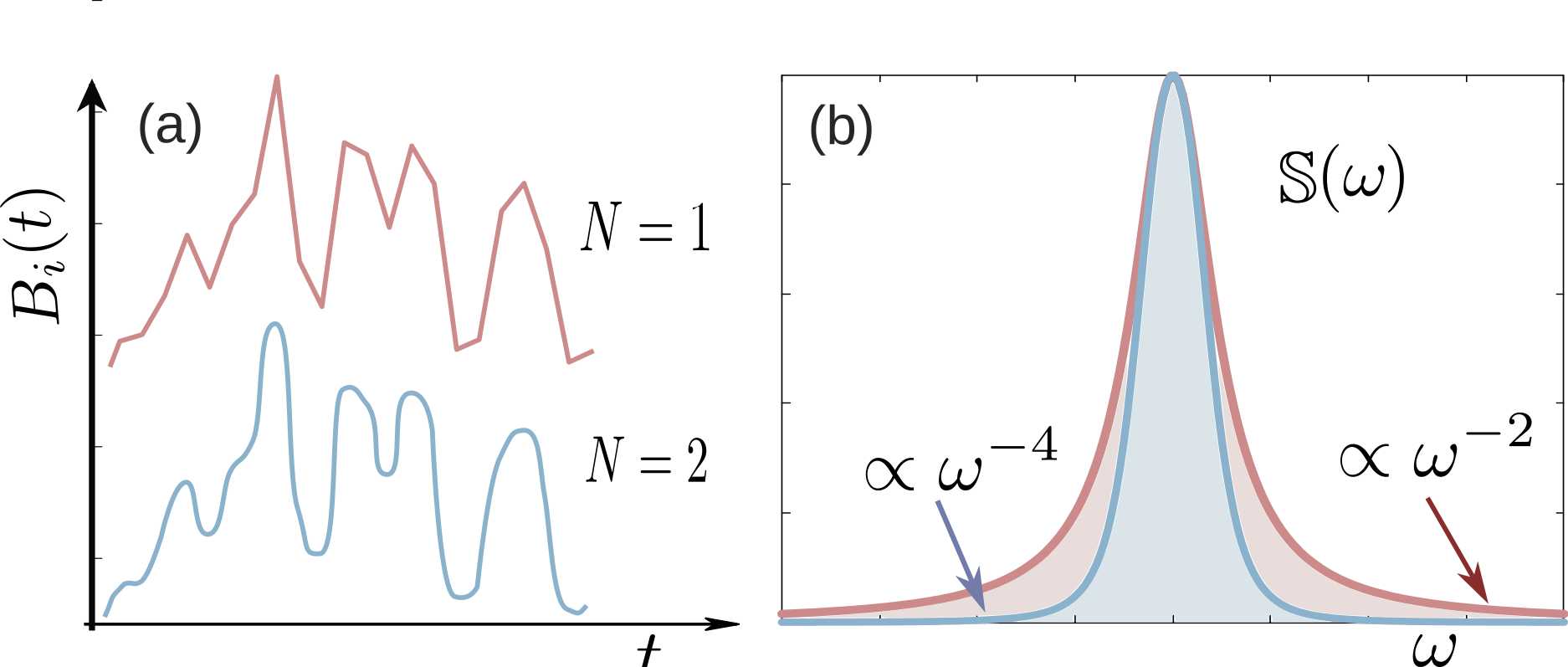}
\par\end{centering}
{\small{}\caption{\label{fig:local-in-time}Schematic comparison between a Markovian
noise for $N=1$ and a non-Markovian noise with $N=2$ for one dimensional,
local-in-time stationary noises with different constraints on the
differential operator $\mathbb{D}$. (a) Examples of single realizations
of field fluctuations $B_{i}(t)$ for both cases. The differential
operator has constraints up to the first derivative for $N=1$, where
the field fluctuation is continuous but not differentiable. For $N=2$,
the field fluctuation and its derivative are continuous. (b) The corresponding
noise spectrums $\mathbb{S}(\omega)$. For $N=1$ the spectral density
is a Lorentzian and for $N=2$ the inverse of a quartic polynomial,
with power law tails $\propto\omega^{-2}$ and $\propto\omega^{-4}$
for large frequencies, respectively.}
}{\small\par}
\end{figure}

The inverse relation of Eq. (\ref{eq:Def_Kernel_G}) for local-in-time
processes becomes
\begin{equation}
\mathbb{D}(t)\mathbb{G}(t,t')=\delta(t-t')\mathbb{I}\label{eq:Def_Local-in-time_G}
\end{equation}
under these assumptions. This relation shows that the time dependent
operator $\mathbb{D}(t)$ can be diagonalized by simply solving a
set of \emph{ordinary differential equations}. This is in contrast
to the case of nonlocal-in-time noise processes, where it is necessary
to solve \emph{integral equations }to obtain the noise eigenmodes.
The autocorrelation functions $\mathbb{G}(t,t')$ are the Green functions
of the local-in-time differential operator $\mathbb{D}(t)$. Therefore
a local-in-time noise process can be purely described by the constraints
$\mathbb{D}_{k}^{H/A}(t)$ on the derivatives of the fluctuating field
$\mathbf{B}(t)$ as seen in Eq. (\ref{eq:local-in-time-action}),
providing more simple physical meanings for the form of the Kernel
operator given in Eq. (\ref{eq:local_in_time_kernel_operator}). A
direct implication of this Kernel form is that it describes fluctuating
field paths that are differentiable up to order \textbf{$N-1$}. This
means that $\mathbf{B}$ and its first \textbf{$N-1$} derivatives
must be continuous functions {[}see Fig. \ref{fig:local-in-time}
for an example{]} so as the action does not diverge and they contribute
to the propagator (See Appendix \ref{subsec:Differentiability-of-the}
for a demonstration). Another implication of this Kernel form is that
its characterization only requires to estimate the matrices $\mathbb{D}_{k}^{H/A}(t)$,
which are  $2N+1$ functions of $\mathfrak{R}\to\mathfrak{R}^{n\times n}$.
Conversely, a general non local-in-time process, requires estimating
the Kernel $\mathbb{D}(t_{1},t_{2})$ which is a function $\mathfrak{R}^{2}\to\mathfrak{R}^{n\times n}$.{\small{}}
{\small\par}

The Kernel operator of these local-in-time fluctuating field processes
is described in the frequency basis as
\begin{multline}
\mathbb{D}(\omega_{1},\omega_{2})=(\omega_{1}|\mathbb{D}|\omega_{2})=\sum_{k=0}^{N}\omega_{1}^{k}\omega_{2}^{k}\mathbb{D}_{k}^{H}(\omega_{1}-\omega_{2})\,\\
+\frac{1}{2}\sum_{k=1}^{N}i\left(\omega_{1}^{k}\omega_{2}^{k-1}+\omega_{1}^{k-1}\omega_{2}^{k}\right)\mathbb{D}_{k}^{A}(\omega_{1}-\omega_{2}),\label{eq:localintimeKernel-frequency_basis}
\end{multline}
where $\mathbb{D}_{k}^{H/A}(\omega_{1}-\omega_{2})=\int\frac{dt}{2\pi}\mathbb{D}_{k}^{H/A}(t)e^{it(\omega_{1}-\omega_{2})}$
are the Fourier transforms of the corresponding time dependent constraints
$\mathbb{D}_{k}^{H/A}(t)$ on the $k$-th derivatives of the field
fluctuations. Then, according to Eq. (\ref{eq:identity_relation_non_stationary}),
its inverse gives the bispectrum $\mathbb{S}(\omega_{1},\omega_{2})$.
Based on Eq. (\ref{eq:localintimeKernel-frequency_basis}), this spectrum
is similar to a polinomial of $\omega_{1}$ and $\omega_{2}$, except
that in the non-stationary case, the ``coefficients'' $\mathbb{D}_{k}^{H/A}(\omega_{1}-\omega_{2})$
depend on the frecuencies, and are determined by the temporal dependence
of the constraints on the the $k$-th derivatives of the field fluctuations.
The non-stationary evolution of the constraints is manifested for
these local-in-time processes by functions that satisfy $\mathbb{D}(\omega_{1},\omega_{2})=\mathbb{D}(\omega_{2},\omega_{1})^{\dagger}$,
and therefore have reflection-conjugation symmetry about the diagonal
axis $\omega_{1}=\omega_{2}$ in the frequency space. In stationary
processes as we see below, $\mathbb{D}_{k}^{H/A}(\omega_{1}-\omega_{2})$
become proportional to a $\delta(\omega_{1}-\omega_{2})$ function,
thus the width of these functions $\mathbb{D}_{k}^{H/A}(\omega_{1}-\omega_{2})$
show the degree of non-stationarity of a process. 

An alternative interpretation of the physical properties of these
processes comes from employing a time discretization of the time evolution
(see Appendix \ref{sec:Discretization-local-in-time}). The argument
of the integral of Eq. (\ref{eq:local-in-time-action}) depends only
on the local time $t$, i.e. the Kernel only correlates the fluctuating
field values and its derivatives instantaneously. However, the correlations
imposed by the Kernel on the field derivatives effectively correlate
the field at different times locally. Replacing the field derivatives
in Eq. (\ref{eq:local-in-time-action}) with finite time diferences,
the discrete version of the Kernel correlates the value of the fluctuating
field at the nearest $2N+1$ times steps close to the time $t$, i.e.
$\mathbf{B}(t\pm k\Delta t)$ with $k=-N...N$ and $\Delta t$ the
time discretization step. Thus, this shows how a kernel operator that
is local-in-time, correlates the fluctuating field values on a limited
correlation time length, introducing therefore memory of the fluctuating
field process $\mathbf{B}(t)$ on infinitesimally near times determined
by the the contraints on the field derivatives. These processes are
in general non-Markovian as discussed in Sec. \ref{sec:Conditions-Markovian},
yet, the long-time correlations do not exits.

\subsection{Stationary noise spectrum as the inverse of the differential operator}

For stationary noises that are local-in-time, the differential operator
$\mathbb{D}$ is time independent
\begin{equation}
\mathbb{D}=\sum_{k=0}^{N}(-1)^{k}\mathbb{D}_{k}^{H}\,\partial_{t}^{2k}+\sum_{k=1}^{N}(-1)^{k}\mathbb{D}_{k}^{A}\partial_{t}^{\,2k-1},
\end{equation}
where we used $\overset{\leftarrow}{\partial_{t}^{\,}}=-\partial_{t}$
(See Appendix \ref{sec:Simplification-Expression-D}). The path integral
framework allows then to express the noise spectrum $\mathbb{S}(\omega)$
in terms of the operators $\mathbb{D}_{k}^{H,A}$ that define the
probability density of field paths. Notice that $\mathbb{D}_{k}^{H,A}$
define now constant constraints on the derivatives of the fluctuating
field $\mathbf{B}(t)$. Figure \ref{fig:local-in-time}(a) shows typical
field fluctuations with constraints up to the derivative of the field
($N=1$) and up to the second derivative of the field ($N=2$). Based
on Eq. (\ref{eq:spectral_density_stationary}), these constraints
then define the noise spectrum as 
\begin{equation}
\mathbb{S}(\omega)=\left[\mathbb{D}(\omega)\right]^{-1}=\left[\sum_{k=0}^{N}\mathbb{D}_{k}^{H}\omega^{2k}+i\sum_{k=1}^{N}\mathbb{D}_{k}^{A}\omega^{\,2k-1}\right]^{-1},\label{eq:S(w)-3}
\end{equation}
where this matrix polynomial $\mathbb{D}(\omega)$ is the expression
in the frequency-basis of the differential operator $\mathbb{D}$
of a local-in-time stationary process. This equation is obtained from
Eq. (\ref{eq:localintimeKernel-frequency_basis}), where $\mathbb{D}_{k}^{H/A}(\omega_{1}-\omega_{2})\propto\delta(\omega_{1}-\omega_{2})$
for stationary processes. The polynomic functional behavior of the
predicted noise spectrum is consistent with several experimental observations
\citep{Lange2010,Alvarez2011,Medford2012,Malinowski2017a}. As examples,
a\emph{ white noise }correspond to \emph{$N=0$}, a\emph{ Lorentzian}
spectrum corresponds to $N=1$ representing Ornstein-Uhlenbeck one
dimensional noise processes, and an spectrum given by the inverse
of a quartic polynomial corresponds to $N=2$ with a power law tail
$\propto\omega^{-4}$ {[}Fig. \ref{fig:local-in-time}(b){]}. This
result thus provides a simple physical meaning for the power law exponents
dependence of the noise spectrum based on the conditions $\mathbb{D}_{k}^{H}$
and $\mathbb{D}_{k}^{A}$ imposed to the derivatives of $\mathbf{B}(t)$
that the probability density of the possible paths can take.

An important application of Eq. (\ref{eq:S(w)-3}) is that it facilitates
the noise spectroscopy characterization, as the noise spectrum of
local-in-time processes is defined by the $2N+1$ \emph{constant }$n\times n$-matrices
$\mathbb{D}_{k}^{H/A}$. Therefore, the hypothesis of locality-in-time
allows the implementation of \emph{parametric estimation} to determine
noise spectra, rather than using non-parametric estimation as described
in Sec. \ref{subsec:Dynamical-decoupling-noise} for more general
noises. Therefore noise spectroscopy methods for local-in-time stationary
noises can thus reduce the amount of measurements required to reconstruct
the noise spectrum. 

\subsection{Conditions for Markovian and non-Markovian noise processes\label{sec:Conditions-Markovian}}

In this subsection, we state the conditions for the kernel operator
$\mathbb{D}$ to represent a Markovian process. We show that non-Markovian,
local-in-time processes can always be mapped to a Markovian process,
which is described by the value of the field and its first $N-1$
derivatives. We show that local-in-time processes are a generalization
of Markovian processes and that those processes cannot generate long-time
correlations. The probability distribution of $\mathbf{B}$ at a time
$t'$ after a time $t$ is shown to be based only on the state of
the process at time $t$, if we consider the generalized process defined
by the state of the joint probability distribution of $\mathbf{B}$
and its first $N-1$ derivatives at time $t$. Therefore, local-in-time
noises are generally non-Marovian and in our generalized framework
appear as a natural extension of Markovian noises, as they maintain
many of their properties while being able to model a greater variety
of environments. They thus provide a natural model to study non-Markovian
environments that does not contain long-time correlations. 

\subsubsection{Markovian process conditions}

A stochastic process $\mathbf{B}$ is Markovian if the probability
of the field $\mathbf{B}_{f}$ at time $t_{f}$ starting from a value
$\mathbf{B}_{0}$ at $t_{0}$ is \citep{breuer_open,Rivas2014,Breuer2016,Milz2021}

{\small{}
\begin{equation}
P(\mathbf{B}_{f},t_{f}|\mathbf{B}_{0},t_{0})=\int\!d\mathbf{B}_{1}\,P(\mathbf{B}_{f},t_{f}|\mathbf{B}_{1},t_{1})P(\mathbf{B}_{1},t_{1}|\mathbf{B}_{0},t_{0}),\label{eq:prop_Markov}
\end{equation}
}for any $t_{f}>t_{0}$, where $P(\mathbf{B}_{i},t_{i}|\mathbf{B}_{j},t_{j})$
are the conditional probabilities that define the propagator of the
process. A Markovian process is said to be memoryless, as the random
variable ``forgets'' its previous state as it evolves \citep{breuer_open,Rivas2014,Breuer2016,Milz2021}.
Therefore, in general a stochastic process determined by the Kernel
operator $\mathbb{D}$ as described in Eq. (\ref{eq:def_gral_accion}),
is non-Markovian as it correlates the fluctuating field states at
two distant times. Only local-in-time noise processes as defined in
Sec. \ref{subsec:Local-in-time-framework} may be Markovian. We then
only consider this case to evaluate the conditions required for the
non-stationary noise process to be Markovian.

If the action $\mathcal{A}[\boldsymbol{\mathbf{B}}]$ of Eq. (\ref{eq:local-in-time-action})
contains derivatives of order $N$ higher than $1$, it implies a
differentiability condition for the field paths so as the action does
not diverge and they contribute to the propagator (See Appendix \ref{subsec:Differentiability-of-the}).
Therefore in this case $\mathbf{B}$ and its first \textbf{$N-1$}
derivatives must be continuous functions. As Eq. (\ref{eq:prop_Markov})
only imposes that the paths are continuous at $t_{1}$ to define a
finite integral, if the \emph{action} imposes more strict conditions
on the possible paths, requiring \emph{continuity} from their \emph{derivatives},
the process of $\mathbf{B}$ will not be Markovian as the propagator
requires information of different previous times. This implies in
particular that the derivative of $\mathbf{B}$ is continuous, and
therefore it contains information about the history of $\mathbf{B}$.
Then Eq. (\ref{eq:prop_Markov}) cannot be satisfied as it does not
include information about $\mathbf{B}$ encoded in its derivatives.
Therefore, we found that the stochastic process that describes \emph{$\mathbf{B}$}
is Markovian if and only if its \emph{Kernel operator} is given by
the \emph{local-in-time differential operator} of the form\emph{ 
\begin{equation}
\mathbb{D}_{\text{Mark}}(t)=\mathbb{D}_{0}^{H}(t)+\overset{\leftarrow}{\partial_{t}}\mathbb{D}_{1}^{H}(t)\,\partial_{t}+\frac{1}{2}\left[\overset{\leftarrow}{\partial_{t}}\mathbb{D}_{1}^{A}(t)-\mathbb{D}_{1}^{A}(t)\partial_{t}\right],\label{eq:Condicion_Markov_sobre_D}
\end{equation}
}where the derivatives of the paths are not continuous\emph{ }(see
Appendix \ref{sec:Non-markovian-noise-demonstratio} for a full proof)\emph{.}
 In the frequency basis, the Kernel becomes
\begin{multline}
\mathbb{D}_{\mathrm{Mark}}(\omega_{1},\omega_{'2})=(\omega_{1}|\mathbb{\mathbb{D}_{\mathrm{Mark}}}|\omega_{2})\\
=\mathbb{D}_{0}^{H}(\omega_{1}-\omega_{2})+\omega_{1}\,\omega_{2}\mathbb{D}_{1}^{H}(\omega_{1}-\omega_{2})\\
+\frac{1}{2}i\left(\omega_{1}+\omega_{2}\right)\mathbb{D}_{1}^{A}(\omega_{1}-\omega_{2}),\label{eq:localintimeKernel-markovian-frequeny}
\end{multline}
following Eq. (\ref{eq:localintimeKernel-frequency_basis}). Equations
(\ref{eq:Condicion_Markov_sobre_D}) and (\ref{eq:localintimeKernel-markovian-frequeny})
are definitions of Markovian Gaussian processes within the presented
framework that are equivalent to the conventional form of Eq. (\ref{eq:prop_Markov})\emph{.}

In the particular case of stationary noises, the most general $n$-dimensional
Markovian noise spectrum is therefore
\begin{equation}
\mathbb{S}_{\text{Mark}}(\omega)=\left[\mathbb{D}_{0}^{H}+i\mathbb{D}_{1}^{A}\omega+\mathbb{D}_{1}^{H}\omega^{2}\right]^{-1},\label{eq:noise_spectrum_markovian_1D}
\end{equation}
according to Ec. (\ref{eq:S(w)-3}). For the simplest case of one
dimensional stationary noises, this means that the process that describes
them\emph{ }is Markovian if and only if the noise spectrum is constant,
i.e. white noise\emph{ }{[}\emph{$N=0$} in Eq. (\ref{eq:S(w)-3}){]},
or Lorentzian {[}$N=1$ in Eq. (\ref{eq:S(w)-3}){]}. Figure \ref{fig:local-in-time}
compares a Markovian field fluctuation for $N=1$, where the field
is not differentiable, and a non-Markovian field fluctuation with
a continuous derivative for a one dimensional stationary noise.

An important application of the noise spectrum form derived in Eqs.
(\ref{eq:localintimeKernel-frequency_basis}) and (\ref{eq:noise_spectrum_markovian_1D})
is that they allow to determine when a noise process is Markovian
or non-Markovian by experimentally measuring its spectrum with dynamical
decoupling noise spectroscopy (Sec. \ref{subsec:Dynamical-decoupling-noise}).

\subsubsection{Generalized Markovian process for local-in-time non-stationary, non-Markovian
noises}

If the stochastic process is non-Markovian, but local-in-time, it
can still be described by a generalized Markovian process if all the
derivatives of $\mathbf{B}$ are included in the stochastic process.
We define a general conditional probability, or propagator, of the
stochastic process based on the ordered set of all the derivatives
of $\mathbf{B}$ of order lower than $N$, 
\begin{equation}
\begin{array}{cc}
\{\mathbf{B}^{(k)}(t)\} & \equiv\{\mathbf{B}^{(k)}(t)\,:\,k=0,\dots N-1\}\\
 & =\{\mathbf{B}(t),\dot{\mathbf{B}}(t),\dots,\mathbf{B}^{(N-1)}(t)\}.
\end{array}
\end{equation}
The Markovian condition of Eq. (\ref{eq:prop_Markov}) is now satisfied
for the stochastic process of $\{\mathbf{B}^{(k)}(t)\}$, as the equation
imposes continuity on the first $N-1$ derivatives of $\mathbf{B}$.
This means that both the action and the integral over $\{\mathbf{B}_{1}^{(k)}(t)\}$
impose that $\{\mathbf{B}^{(k)}(t)\}$ are continuous, implying that
the \emph{stochastic process $\{\mathbf{B}^{(k)}(t)\}$ is Markovian}
(see the proof details in Appendix \ref{sec:Markovian-full-stochastic}).
Therefore the family of \emph{fluctuating local-in-time non-stationary
noises} considered here imply that if $\mathbf{B}$ is not Markovian,
the state of $\mathbf{B}$ is not only determined by the \emph{probability
distribution} of $\mathbf{B}$, but by the \emph{joint probability
distribution} of $\mathbf{B}$ and its derivatives. This means
that the information of the initial condition is encoded in the \emph{derivatives}
of $\mathbf{B}$.

Since local-in-time noises can be described by the generalized Markovian
process of $\{\mathbf{B}^{(k)}(t)\}$, its \emph{propagator} $P(\{\mathbf{B}_{f}^{(k)}\},t_{f}|\{\mathbf{B}_{0}^{(k)}\},t_{0})$
contains all the \emph{information} about the stochastic process that
describes the field paths $\mathbf{B}$. Moreover, we found that the
path integral framework allows to calculate the \emph{propagator}
of $\{\mathbf{B}^{(k)}\}$ without actually requiring to perform \emph{path
integrals}. Instead, the propagator can be obtained by solving an
\emph{ordinary linear differential equation }with different types
of \emph{boundary conditions. }The full propagator expression and
its derivation are given in the Appendix \ref{sec:Markovian-Propagator}.

In summary, the local-in-time stochastic process of $\mathbf{B}$
is Markovian if and only if the differential operator $\mathbb{D}(t)$
does not contain terms with $k$ higher than $1$. All other processes
are non-Markovian. Yet, the generalized local-in-time process of $\{\mathbf{B}^{(k)}(t)\}$
is always Markovian, and it follows that the information of previous
states of $\mathbf{B}$ is encoded in its derivatives when they are
continuous. 

\subsubsection{The long time limit of local-in-time stationary noises is indistinguishable
from a Markovian process}

The spectral density $\mathbb{S}(\omega)$ for local-in-time stationary
noises of Eq. (\ref{eq:S(w)-3}) is infinitely differentiable, as
it is the inverse of a non-singular matrix polynomial, and therefore
the integral of its derivatives is finite $\int d\omega\,\left|\left(\frac{d}{d\omega}\right)^{k}\mathbb{S}_{ij}(\omega)\right|<\infty$
for all $k$-th derivative order. Since the corresponding correlation
functions are $\mathbb{G}_{ij}(t)=\int\frac{d\omega}{2\pi}\mathbb{S}_{ij}(\omega)e^{-i\omega t}$,
one can see that 
\begin{equation}
\mathbb{G}_{ij}(t)=(it)^{-k}\int\frac{d\omega}{2\pi}\left(\frac{d}{d\omega}\right)^{k}\mathbb{S}_{ij}(\omega)e^{-i\omega t}
\end{equation}
 and that they are bounded by the expression \citep{Soueycatt2016}
\begin{equation}
\left|\mathbb{G}_{ij}(t)\right|\leq\frac{\int\frac{d\omega}{2\pi}\left|\left(\frac{d}{d\omega}\right)^{k}\mathbb{S}_{ij}(\omega)\right|}{\left|t^{k}\right|}.
\end{equation}
Therefore, we obtain that $\lim_{|t|\to\infty}\frac{\mathbb{G}(t)}{t^{k}}=0$
for all $k$, implying that $\mathbb{G}(t)$ decays to zero exponentially
(or faster) for long times $t\to\infty$. Since $\mathbb{S}(\omega)$
is the inverse of a matrix polynomial, we know that it is not entire
when analytically continued to the complex plane of $\omega$ for
$N>0$. Therefore for $N>0$, the correlation functions $\mathbb{G}(t)$
cannot decay faster than an exponential in the long time limit, so
they must decays exactly as exponentials \citep{Reed1972}. Only for
$N=0$, the correlation function decays faster as its corresponding
noise is a white noise.  Therefore in the long time limit, every
local-in-time stationary noise is indistinguishable by means of noise
spectroscopy from a Markovian noise, as its self-correlation functions
decay exponentially or are zero for long times, which coincides with
the correlations functions derived from what we showed are Markovian
noises.

\subsubsection{Short time limit $t\to0$ of local-in-time stationary noises}

We consider an arbitrary differential operator $\mathbb{D}$ for
a stationary local-in-time noise. In general, it is not possible to
find a closed formula for the correlation functions $\mathbb{G}(t)$,
since that would imply finding the roots of a polynomial of an arbitrarily
high degree. However, it is possible to get information about the
general behavior of $\mathbb{G}(t)$ at short times, by analyzing
its representation in the frequency-basis, i.e. the noise spectrum
$\mathbb{S}(\omega)$. 

In the large frequency $\omega$ limit, the noise spectrum is $\mathbb{S}(\omega)=\mathbb{D}^{-1}(\omega)\sim\mathbb{D}_{N}^{-1}\omega^{-2N}$
according to Eq. (\ref{eq:S(w)-3}). Since $\mathbb{D}(\omega)$ is
positive definite, the spectral density $\mathbb{S}(\omega)$ remains
finite for all $\omega$ and the integral $\int d\omega\left|\omega^{2N-2}\mathbb{S}_{ij}(\omega)\right|<\infty$
is bounded. This implies that $\mathbb{G}(t)$ is continuously differentiable
$2N-2$ times, since its derivatives are 
\begin{equation}
\mathbb{G}_{ij}^{(k)}(t)=\int\frac{d\omega}{2\pi}(-i\omega)^{k}\mathbb{S}_{ij}(\omega)e^{-i\omega t}
\end{equation}
and are continuous \citep{Soueycatt2016}. This means that for processes
with $N>1$, the correlation functions decay quadratically or slower
around their maxima. Then as the correlation functions of each component
of $\mathbf{B}$ satisfy 
\begin{equation}
\mathbb{G}_{ii}(t)=\langle B_{i}(t)B_{i}(0)\rangle=\text{Cov}(B_{i}(t)B_{i}(0))\leq\langle B_{i}^{2}(0)\rangle,
\end{equation}
they have a global maximum at $t=0$, implying that if $N>1$ they
decay at least quadratically for short times. This agrees well with
the expected behavior of correlation functions predicted by quantum
mechanics \citep{Misra1977,Kofman2001,Kofman2004a,Facchi2004,Facchi2005,Danieli2005,Alvarez2006,Pascazio2014,Virzi2021}. 

Therefore, every stationary noise process generated by a quantum
mechanical process is not Markovian. Notice that the Markovian noise
correlation functions for $N\leq1$ are a Dirac delta function that
describes a white-noise spectrum ($N=0$), and for a one dimensional
noise with $N=1$, we obtain a Lorentzian spectrum that describes
an \emph{Ornstein-Uhlenbeck process} where $\mathbb{G}(t)=\frac{1}{2D_{0}\tau_{c}}e^{-|t|/\tau_{c}}$,
with $\tau_{c}=\sqrt{\nicefrac{D_{1}}{D_{0}}}$. These Markovian noise
processes disagree with what quantum mechanics predicts, that for
short times $\mathbb{G}$ should decay at least quadratically. This
is true for all Markovian processes, as Eq. (\ref{eq:Def_Local-in-time_G})
demands for the first derivative of $\mathbb{G}(t,t')$ to be discontinuous
at $t=t'$ when $N=1$.

\section{\label{sec:Non-stationary-noises}Implementation of the framework
on paradigmatic non-stationary noises}

In this section, we show how to implement the presented path integral
framework to two paradigmatic examples of non-stationary noises. We
consider a noise determined from a quench on the environment and a
noise that acts near to a point of time. Both examples shows some
characteristic features that arise from our framework that distinguish
non-stationary from stationary noise effects on the qubit-probe dephasing.

\subsection{Quenched diffusion}

We here consider a quenched environment described by a diffusion process
of the fluctuating field that begin at an instant of time . This sets
a paradigmatic model of typical environments that can be suddenly
quenched to put them out of equilibrium, where excitations start spreading
over a large number of degrees of freedom \citep{Polkovnikov2011,Alvarez2015,Eisert2015,Abanin2019,Wang2021}.
The environment dynamics sensed by the qubit-probe can be represented
with a generalized diffusion process, and possible examples can be
encountered in spin ensembles coupled to single NV centers in diamond
\citep{Wang2012,Luan2015,Romach2015}, macromolecular dynamics \citep{Mittermaier2006,SchaeferNolte2014,Staudacher2015,Shi2015},
spin diffusion on environments that becomes out of equilibrium \citep{suter_spin_1985,Alvarez2011,Alvarez2013,Alvarez2015,Alvarez2015a},
dynamics of spin and current fluctuations in a material at the nanoscale
probed by magnetic noise sensors \citep{SchmidLorch2015,Casola2018},
and molecular diffusion out of equilibrium \citep{Steinert2013,Ziem2013,Chakrabarti2016,Cohen2020,Chalermpusitarak2020}.

\begin{figure}
\begin{centering}
\includegraphics[width=0.9\columnwidth]{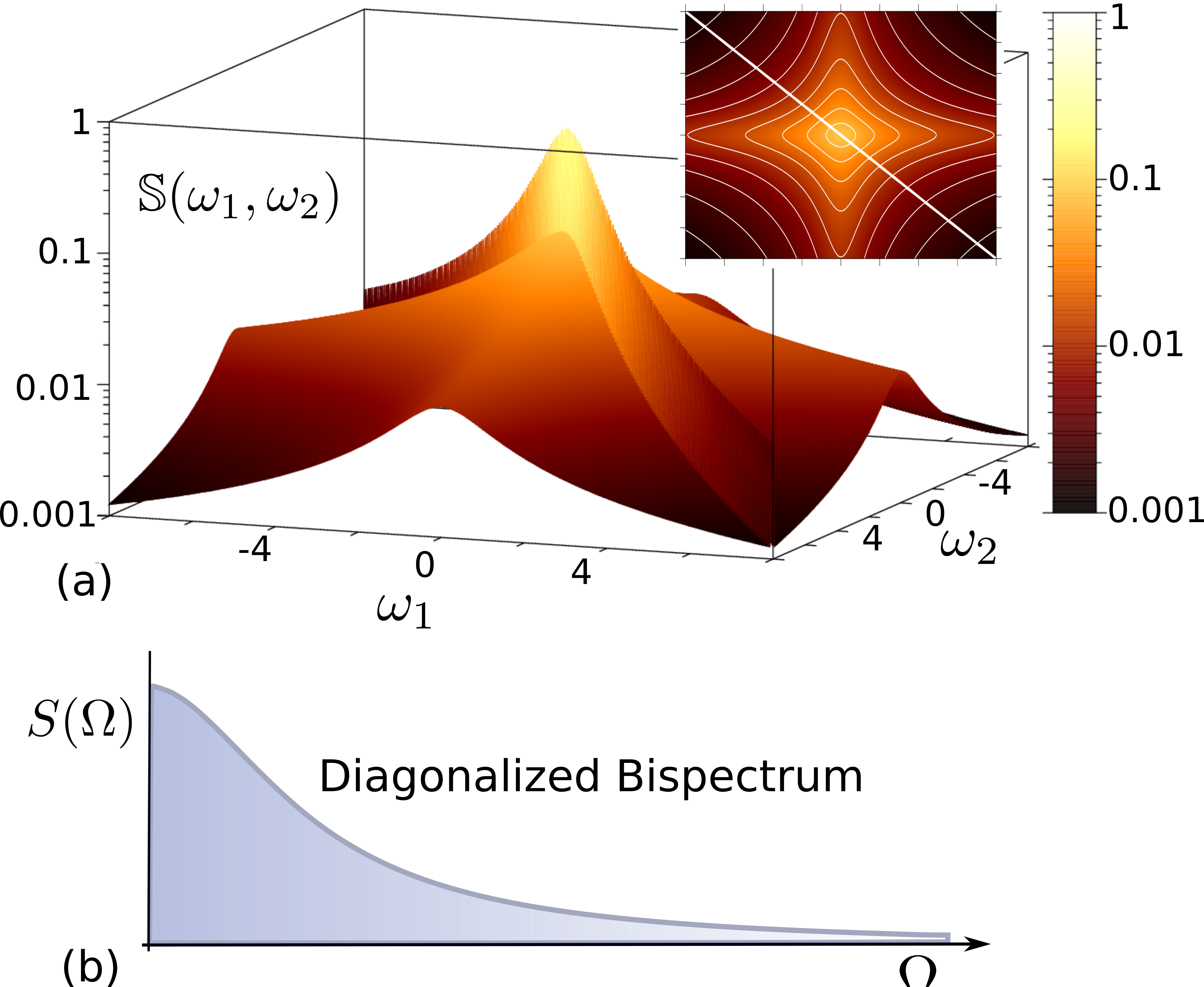}
\par\end{centering}
{\small{}\caption{\label{fig:bispectrum}Non-stationary noise spectrum of a one dimensional
quenched diffusion process of the environmental field fluctuations,
driven by a one dimensional Ornstein-Uhlenbeck stochastic process.
(a) The absolute value of the corresponding spectral density $\mathbb{S}(\omega_{1},\omega_{2})$
in the frequency-basis, the bispectrum given by Eq. (\ref{eq:quenched_diffusion_spectrum}),
is shown. Only the diagonal $\omega_{1}=\omega_{2}$ contains the
stationary component of the noise spectrum that is dominant at long
times after the quench. As the stationary component is $\propto\delta(\omega_{1}-\omega_{2})$,
we have plotted it by replacing the $\delta(\omega_{1}-\omega_{2})$
function by a non-divergent gaussian function for presentation purposes.
The inset shows a surface colored map of the bispectrum with contour
lines in white color. (b) The corresponding generalized, non-stationary
noise spectrum $S(\Omega)$, given in the noise eigenmode basis $\left|\Omega\right)$
that diagonalize the noise spectrum.}
}{\small\par}
\end{figure}

As a simple model, in particular we consider that the fluctuating
field is zero and at a given time suddenly starts to fluctuate, driven
by a one dimensional Ornstein-Uhlenbeck diffusion process \citep{Lange2010,Luan2015,SchmidLorch2015,Romach2015,zwick_maximizing_2016,zwick2020precision}.
Considering that the quench is at time $t=0$, the fluctuating field
is therefore confined to a point for $t<0$. For $t>0$ the noise
process is described by a differential operator equal to a stationary
one that gives a Lorentzian spectrum
\begin{equation}
S_{0}(\omega)=\frac{1}{D_{0}+D_{1}\omega^{2}}.\label{eq:Ornstein_stationary}
\end{equation}
This quenched diffusion noise process is local-in-time and is modeled
by the differential operator 
\begin{equation}
\mathbb{D}(t)=D_{1}\partial_{t}^{2}+D_{0}(t),
\end{equation}
with 
\begin{equation}
D_{0}(t)=\begin{cases}
D_{0} & t>0\\
+\infty & t<0
\end{cases},
\end{equation}
where we considered $\mathbb{D}$ as a scalar Kernel operator. The
action and the differential operator that describe the possible field
paths are equivalent to the ones derived from a Schrödinger equation
that describe an infinite potential wall at the position $0$. Based
on this analogy one can consider that $D_{1}\partial_{t}^{2}$ is
mapped to $-\frac{1}{2m}\partial_{x}^{2}$ and $D_{0}(t)$ is mapped
to the energy potential that describe the wall by replacing $t$ with
the position $x$. Therefore, the eigenmode basis that diagonalizes
the Kernel operator $\mathbb{D}$ is 
\begin{equation}
|\Omega)=\sqrt{\frac{2}{\pi}}\int_{0}^{+\infty}dt\,\sin(\Omega t)|t),\label{eq:noise_eigenmodes-quenched_diffusion}
\end{equation}
where $(\Omega|\Omega')=\delta(\Omega-\Omega')$ and $\Omega\geq0\in\mathbb{R}$.
Following Eq. (\ref{eq:spectral_density_lamda}), the noise spectrum
on the eigenmode basis of the non-stationary field fluctuations is
(Fig. \ref{fig:bispectrum})
\begin{equation}
S(\Omega)=\frac{1}{D_{0}+D_{1}\Omega^{2}},\label{eq:Quenched_diffusion_spectrum}
\end{equation}
and the generalized filter function is defined by
\begin{equation}
F(\Omega)=(\Omega|\boldsymbol{f})=\int_{0}^{+\infty}dt\:\sqrt{\frac{2}{\pi}}\sin(\text{\ensuremath{\Omega}}t)f(t).\label{eq:F_Dirichlet_en_0}
\end{equation}
{\small{}}
Notice how the functional form of the non-stationary noise spectrum
of Eq. (\ref{eq:Quenched_diffusion_spectrum}) is \emph{equal }to
that of a \emph{stationary} case for $\omega\geq0$ {[}Eq. (\ref{eq:Ornstein_stationary}){]}
replacing $\Omega$ by $\text{\ensuremath{\omega}}$. However the
\emph{eigenmodes} are different to the ones obtained in the stationary
case. In this example, this means that the difference between the
stationary and non-stationary cases is encoded in the filter function
$F$. For a stationary Ornstein-Uhlenbeck process, $F$ is obtained
by performing the complex Fourier transform of $f(t)$, but in the
non-stationary process $F$ is obtained by a real Fourier transform
based on the sinusoidal component given by Eq. (\ref{eq:F_Dirichlet_en_0}).

\subsubsection{Manifestation of the quench on the noise correlation functions}

 The correlation function is the solution of the local-in-time
differential equation $\mathbb{D}(t)\mathbb{G}(t,t')=\delta(t-t')$.
The correlation function satisfies $\mathbb{G}(t,t')=\mathbb{G^{\dagger}}(t',t)$,
$G(0,t')=0$ due to the fluctuating field is 0 at $t=0$ and it also
satisfies $\lim_{t\to\infty}\mathbb{G}(t,t')=0$ for a fixed $t'$
as the correlation function of the Ornstein-Uhlenbeck process decays
exponentially to 0. The solution for the differential equation is
thus
\begin{equation}
\mathbb{G}(t,t')=\frac{1}{2D_{0}\tau_{c}}\left[e^{-\left|t-t'\right|/\tau_{c}}-e^{-\left(t+t'\right)/\tau_{c}}\right]\Theta(t)\Theta(t'),\label{eq:correlation_quenched_diffusion}
\end{equation}
where $\tau_{c}=\sqrt{\frac{D_{1}}{D_{0}}}$ is the correlation time
and $\Theta(t)$ is the Heaviside function. 

For long times after the quench $t/\tau_{c},t'/\tau_{c}\gg1$, we
recover the correlation function for a Ornstein-Uhlenbeck stationary
diffusion process 
\begin{equation}
\mathbb{G}(t,t')\simeq\frac{1}{2D_{0}\tau_{c}}e^{-\left|t-t'\right|/\tau_{c}}.
\end{equation}
In this case $\mathbb{G}(t,t')$ is given by the Green function of
the differential Kernel operator $D_{1}\partial_{t}^{2}+D_{0}$ acting
over time dependent functions with $t\in\mathbb{R}$ instead of $t>0$
as in the quenched case. Therefore, for times much longer than \emph{$\tau_{c}$,
}the environment \emph{forgets }the \emph{boundary condition} imposed
by the quench\emph{.} The non-stationary effects induced by the quench
are thus modeled by the second term 
\[
\frac{-1}{2D_{0}\tau_{c}}e^{-\left(t+t'\right)/\tau_{c}}
\]
of Eq. (\ref{eq:correlation_quenched_diffusion}). The quench effects
are thus only manifested for \emph{times much lower than $\tau_{c}$
}with respect to the quench time $t=0$, in contrast with the stationary
effects that depend on the time correlation difference $\left|t-t'\right|$.

The corresponding spectral density in the frequency-basis of Eq. (\ref{eq:dephasing-decay_f-G-freq})
is the bispectrum
\begin{multline}
\mathbb{S}(\omega_{1},\omega_{2})=\delta(\omega_{1}-\omega_{2})\frac{1}{D_{0}}\frac{1}{1+\omega_{1}^{2}\tau_{c}^{2}}\\
-\frac{\tau_{c}}{4\pi D_{0}}\frac{1}{\left(1-i\omega_{1}\tau_{c}\right)\left(1+i\omega_{2}\tau_{c}\right)},\label{eq:quenched_diffusion_spectrum}
\end{multline}
where the first term is the stationary noise spectrum of Eq. (\ref{eq:Ornstein_stationary}),
and the other term is due to the quench. Figure \ref{fig:bispectrum}(a)
shows this non-stationary bispectrum, where the stationary component
is only manifested on the diagonal $\omega_{1}=\omega_{2}$. The noise
eigenmodes {[}Eq. (\ref{eq:noise_eigenmodes-quenched_diffusion}){]}
and the non-stationary noise eigen-spectrum {[}Eq. (\ref{eq:Quenched_diffusion_spectrum}){]}
are obtained by diagonalising this bispectrum. Figure \ref{fig:bispectrum}(b)
shows the corresponding diagonalized non-stationary spectrum $S(\Omega)$
of Eq. (\ref{eq:Quenched_diffusion_spectrum}).

\subsubsection{Manifestation of the quench on the qubit-probe dephasing}

\emph{}The difference in the qubit-probe decay induced by this quenched
diffusion, between the non-stationary effects and its stationary counterpart
is given by 
\begin{equation}
\begin{array}{c}
\frac{-1}{2D_{0}\tau_{c}}\int dt_{1}dt_{2}\,\mathbf{f}(t)e^{-\left(t+t'\right)/\tau_{c}}\mathbf{f}(t')=\\
=\frac{-1}{2D_{0}\tau_{c}}\left[\int dt\,\mathbf{f}(t)e^{-\nicefrac{t}{\tau_{c}}}\right]^{2}.
\end{array}
\end{equation}
Therefore the non-stationary effects for this type of quenched noise
always \emph{reduce} the dephasing induced on the qubit-probe compared
with a stationary process. We consider in particular the control modulation
function of the typical continuous wave irradiation to show this manifestation
of the non-stationary effects on the qubit-probe dephasing. We start
the control at time $t_{0}$ after the quench 
\begin{equation}
\mathbf{f}(t)=g\cos[\omega(t-t_{0})]\Theta(t-t_{0})\Theta(t_{0}+T-t),
\end{equation}
and it acts during a time $T$. The argument of the dephasing given
in Eq. (\ref{eq:dephasing-decay_f-G}) at time $t_{0}+T$ is
\begin{multline}
(\mathbf{f}|\mathbb{G}|\mathbf{f})=(\mathbf{f}|\mathbb{G}_{0}|\mathbf{f})\\
-\frac{g^{2}\tau_{c}}{2D_{0}}\left\{ \frac{1+[\omega\tau_{c}\sin(\omega T)-\cos(\omega T)]e^{-T/\tau_{c}}}{1+\omega^{2}\tau_{c}^{2}}\right\} ^{2}e^{-2t_{0}/\tau_{c}},
\end{multline}
where $(\mathbf{f}|\mathbb{G}_{0}|\mathbf{f})$ is the dephasing obtained
for the stationary case, and the second term is due to the quench
--non-stationary-- effects. This second term provides a dephasing
term that oscillates with the control frequency $\omega$ as a function
of the duration of the control modulation $T$. This oscillation is
attenuated with the exponential decay $e^{-T/\tau_{c}}$, thus it
disappears at long times after the quench. Then, if the control modulation
duration is $T\gg\tau_{c}$, we still have an extra term due to the
quench effects
\begin{equation}
(\mathbf{f}|\mathbb{G}|\mathbf{f})=(\mathbf{f}|\mathbb{G}_{0}|\mathbf{f})-\frac{g^{2}\tau_{c}}{2D_{0}}\frac{e^{-2t_{0}/\tau_{c}}}{\left(1+\omega^{2}\tau_{c}^{2}\right)^{2}}.\label{eq:dephasing+shift}
\end{equation}
Notice that this last term provides a constant term into the dephasing
that contains information about the quench if $t_{0}\lesssim\tau_{c}$
as it decays exponentially with the time $t_{0}$ when qubit-probe
control started. This prediction thus gives a tool to probe the quench
effects by monitoring the dephasing change as a function of $t_{0}$,
and gives a method to probe the self-correlation times induced by
the quench. Moreover, if one monitor the decay rate of the qubit-probe
as typically done in several noise spectroscopy approaches \citep{Alvarez2011,Almog2011,Bylander2011},
one would obtain only the stationary decay rate given by 
\begin{equation}
(\mathbf{f}|\mathbb{G}_{0}|\mathbf{f})\propto S_{0}(\omega)T.
\end{equation}
Therefore, the effects of the quench are not manifested on the decay
rate, but they are evidenced on a \emph{shift} of the decaying signal
given by the second term of Eq. (\ref{eq:dephasing+shift}).

\subsection{Noise produced near to a point of time}

As one of our results is that non-stationary spectrums can be discrete
according to Eq. (\ref{eq:spectral_density_lamda}), we exploit an
analogy with the Schrödinger equation to describe a paradigmatic example
that manifests this discrete nature. A noise that contains these discrete
features, is for example one that acts only near to an instant of
time, a pulsed noise interaction. This is a paradigmatic model for
a quantum probe that interacts with a noise during a finite duration
of time. Examples of this can be moving charges or particles that
pass near to quantum sensor \citep{Hall2010,Ziem2013,Steinert2013,Kaufmann2013,Hall2013,Tetienne2017},
forces or interactions detected by a moving cantilever or tip that
contains the sensor \citep{Luan2015,Simpson2016,Poggio2018}, and
biomedical applications as the detection of neuronal activity \citep{Hall2012,Hall2013,Zhang2021a}.

Every non-stationary noise described by a differential operator of
the form 
\begin{equation}
\mathbb{D}(t)=-D_{1}\partial_{t}^{2}+\mathbb{D}_{0}(t),
\end{equation}
where $D_{1}$ is a scalar, can be mapped to a quantum-mechanical
problem with the Hamiltonian 
\begin{equation}
\mathcal{H}=-\frac{1}{2m}\partial_{x}^{2}+V(x)
\end{equation}
by replacing $x\to t$, $m\to\frac{1}{2D_{1}}$ and $V(x)\to\mathbb{D}_{0}(t)$.
Therefore for every solvable quantum-mechanical Hamiltonian with positive
eigenvalues, we obtain a solution for the noise spectrum and its eigenmodes
of a Markovian local-in-time, non-stationary Gaussian noise. The energy
levels of the Hamiltonian must be positive so as the differential
operator $\mathbb{D}(t)$ is positive definite, but every bounded
from below Hamiltonian can be transformed into a positive definite
one by adding a large enough constant $C$ to $\mathcal{H}\to\mathcal{H}+C$.

\begin{figure}
\begin{centering}
\includegraphics[width=0.8\columnwidth]{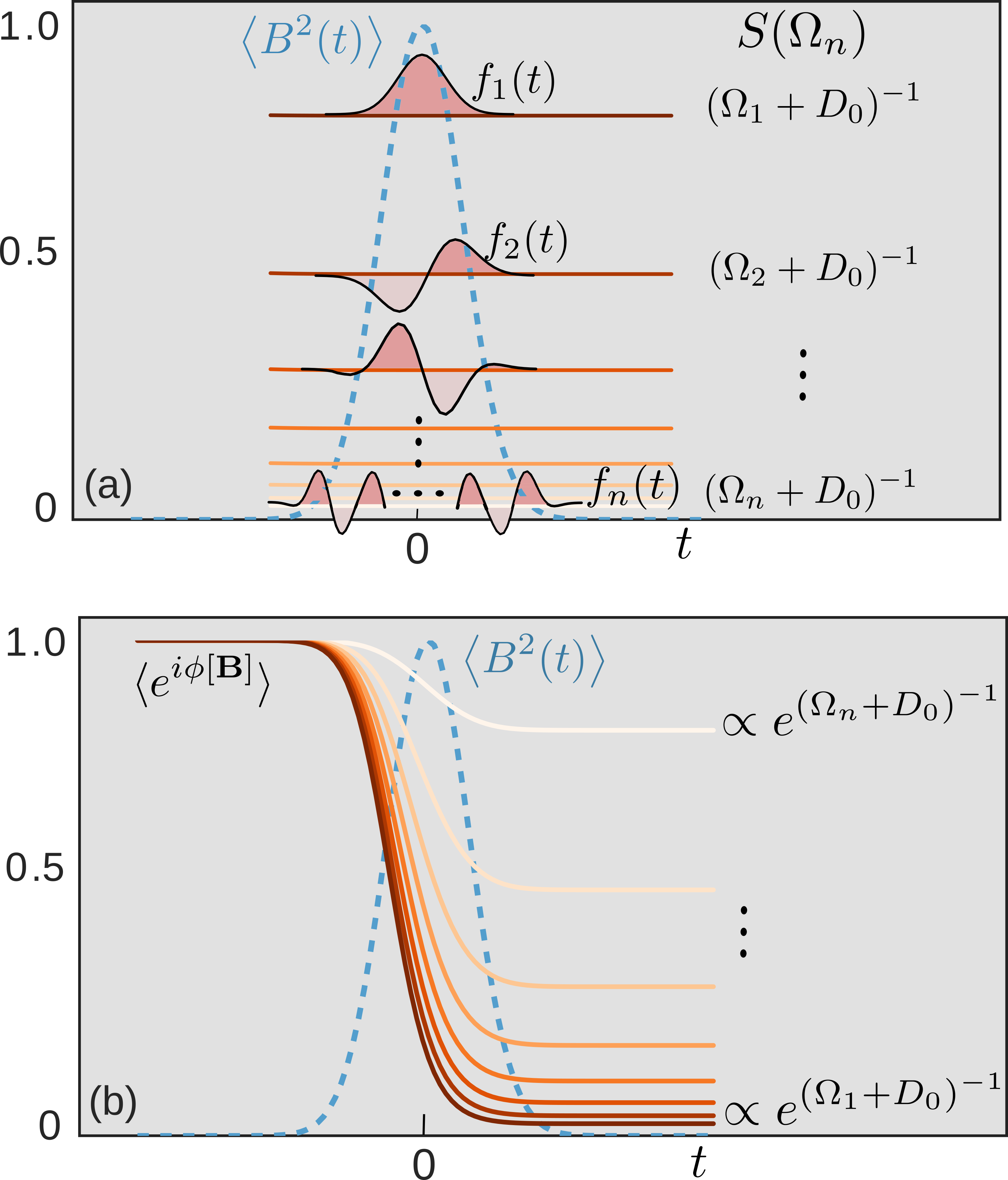}
\par\end{centering}
{\small{}\caption{\label{fig:Qharmosc}Scheme for the characteristic features of a paradigmatic
local-in-time, non-stationary noise that acts only near to a point
of time and therefore produce a discrete, generalized noise spectrum.
The variance of a one dimensional noise that appears near to a point
of time is shown with a blue dashed line. (a) The generalized non-stationary
noise spectrum $S(\Omega_{n})=\frac{1}{\Omega_{n}+D_{0}}$ in the
noise eigenmode basis is shown with orange horizontal lines. The corresponding
noise eigenmodes $|\Omega_{n})$ are shown in the time-basis $f_{n}(t)=(t|\Omega_{n})$
for $n=0,1,2,...$ with solid black lines. They also provide the natural
control modulations to probe the noise eigenmodes. (b) Schematic representation
of the qubit-probe signal decay when it is controlled by the modulation
functions $f_{n}(t)$. They feature a signal saturation at long times
given by $\exp\left\{ -(\Omega_{n}|\mathbb{G}|\Omega_{n})\right\} =\exp\left\{ -S(\Omega_{n})\right\} $.}
}{\small\par}
\end{figure}

Based on this analogy, we describe the paradigmatic example of a pulsed
noise with a one dimensional, non-stationary noise process that is
local-in-time. We consider the noise described by a differential operator
$\mathbb{D}$ that is mappable to the Hamiltonian of a quantum harmonic
oscillator
\begin{equation}
\mathcal{H}=-\frac{1}{2m}\partial_{x}^{2}+\frac{1}{2}m\omega_{0}^{2}x^{2}+D_{0},
\end{equation}
where $D_{0}$ is an additive constant that does not change the eigenvectors
of the Hamiltonian $\mathcal{H}$. The corresponding differential
operator is 
\begin{equation}
\mathbb{D}(t)=-D_{1}\partial_{t}^{2}+D_{0}+\alpha t^{2},
\end{equation}
where the necessary map is $m\rightarrow\frac{1}{2D_{1}}$, $\omega_{0}\rightarrow\sqrt{4\alpha D_{1}}$
and $x\rightarrow t$. This stochastic process models a noise probed
by the qubit-system that appears near to a point of time, where the
fluctuating field paths are forced to be $0$ for times $|t|\to\infty$,
since $D_{0}(|t|\to\infty)\rightarrow\infty$. Therefore, the fluctuating
fields are only allowed to deviate from $0$ near to the local instant
of time $t=0$ (see blue dashed line in Fig. \ref{fig:Qharmosc}).{\small{}}
 The noise eigenmodes of $\mathbb{D}(t)$ are 
\begin{multline}
|\Omega_{n})=\frac{1}{\sqrt{2^{n}n!}}\left(\sqrt{\frac{\alpha}{D_{1}\pi^{2}}}\right)^{1/4}\times\\
\int_{-\infty}^{+\infty}dt\,\exp\left\{ -\sqrt{\frac{\alpha}{4D_{1}}}t^{2}\right\} \,H_{n}\left\{ \left(\frac{\alpha}{D_{1}}\right)^{1/4}t\right\} |t),
\end{multline}
where $H_{n}$ are the Hermite polynomials with $n=0,1,2,...$ in
analogy with the eigenfunctions of the quantum harmonic oscillator
{[}Fig. \ref{fig:Qharmosc}(a){]}. The non-stationary noise spectrum
expressed on its eigenmode basis is then discrete, given by
\begin{equation}
S(\Omega_{n})=\frac{1}{\Omega_{n}+D_{0}},\:\Omega_{n}=\left(n+\frac{1}{2}\right)\omega_{0},
\end{equation}
as shown in Fig. \ref{fig:Qharmosc}(a). Since the differential operator
$\mathbb{D}(t)$ is positive definite if and only if all its eigenvalues
are positive, the model is well defined only for $\omega_{0}>-2D_{0}$.

This noise model sets a paradigmatic example of some of the differences
between probing non-stationary and stationary noise spectra with dynamical
decoupling noise spectroscopy. In this case, the natural control modulation
would be $f_{n}(t)=(t|\Omega_{n})$ for $n=0,1,2,...$ as shown in
Fig. \ref{fig:Qharmosc}(a). Here the dephasing in the long time
limit after applying these modulations will saturate to a constant
value that provides the noise eigenvalues
\begin{equation}
(\Omega_{n}|\mathbb{G}|\Omega_{n})=\frac{1}{\Omega_{n}+D_{0}}.
\end{equation}
This is in contrast to the predicted exponential decay with a constant
rate for stationary noises typically used for noise spectroscopy \citep{Alvarez2011}.
Figure \ref{fig:Qharmosc}(b) shows a schematic representation of
the qubit-probe signal decay when it is controlled by the modulation
functions $f_{n}(t)=(t|\Omega_{n})$. This behaviour is universal
for all noises with a discrete spectrum, therefore this example shows
how noise spectroscopy must be done for such noises.

\section{Summary and Conclusions\label{sec:Summary-and-Conclusions}}

We introduced a path integral framework for determining the dephasing
on a quantum probe induced by non-stationary Gaussian noises. This
type of noises models a fluctuating qubit-probe interaction with a
quantum environment that is out-of-equilibrium under the so-called
weak coupling approximation. We show that the noise generated by this
non-stationary environment is completely determined by a bispectrum
defined by the inverse of the Kernel operator that describes the probability
density of the field paths. This complement recent results, where
similar bispectra functions characterize stationary noises that are
non-Gaussian \citep{Norris2016,Sung2019}. 

The presented framework introduced a generalized noise spectrum for
non-stationary environments, defined by the inverse of the eigenvalues
of the kernel operator that determines the probability of the noise
field paths. The noise eigenmodes define the proper basis to generalize
a filter function derived from the control, whose overlap with the
noise spectrum determines the qubit-probe dephasing. This results
into an extension of the validity of the universal formula for the
dephasing of open quantum systems that depends on the overlap between
a noise spectral density and a qubit-control filter function. The
main result of this generalization is that it allows to implement
two important tools, already developed for stationary noises, to probe
spectral properties and to mitigate decoherence effects of non-stationary
environments. 

We then considered a broad subclass of non-stationary noises, that
we called local-in-time. We show they are described by a differential
operator based on constraints to the derivatives of the fluctuating
field paths. We also show how these constraints are reflected on
the functional behavior of the non-stationary noise spectrums. Such
subclass of the discussed non-stationary noises are the only ones
that can be Markovian if the first derivative of the fluctuating fields
is not continuous, e.g. white noises and stationary noises with Lorentzian
noise spectra. Therefore, if the field derivative is continuous or
the noise is not local-in-time, the noise process is non-Markovian.
Remarkably, we found that local-in-time, non-stationary Gaussian noises
that are non-Markovian, still can be described by a generalized Markovian
noise process that includes the field and a finite number of its derivatives.
This approach simplifies the noise description as it is fully determined
by a differential equation and the constraints on its derivatives
thus redusing the dimensionality for the characterization of the noise
spectra. An important application for the derived forms of the
noise spectrum is that they allow to determine when a noise process
is Markovian or non-Markovian by experimentally measuring its spectrum
with dynamical decoupling noise spectroscopy. In the particular case
of stationary noises that are local-in-time, we found that the differential
equation gives a noise spectra given by the inverse of matrix polynomials
of the frequency modes. The polynomial coefficients associated to
a power of the frequency are given by the constraint to the corresponding
order of the fluctuating field derivative.  This thus allow the implementation
of parametric estimation methods to determine the noise spectra, rather
than using non-parametric estimation that is much more complex and
requires more available experimental time.

We have shown that in some cases local-in-time, non-stationary noises
can be mapped to the Schrödinger equation. Thus every \emph{solvable
1 dimensional quantum mechanical} Hamiltonian $H$ for a single particle
in one spatial dimension with eigenvalues bounded from below, generates
a whole class of solvable noise probability distributions. We have
used this map to apply the presented path integral framework to two
paradigmatic non-stationary noises: a noise acting near to a point
of time --a pulsed noise-- by analogy to a quantum harmonic oscillator
and a quench on the environment that suddenly starts an Ornstein-Uhlenbeck
diffusion process. Both of these examples describe out-of-equilibrium
environments. The pulsed noise is a paradigmatic model for a quantum
probe interacting with a noise during a finite duration of time that
manifests a discrete nature on the generalized non-stationary noise
spectrum. In this case the noise spectrum is obtained by saturation
of the qubit-probe dephasing rather than on its decay rate. The quenched
diffusion noise sets a paradigmatic model for typical environments
that can be suddenly quenched to put them out of equilibrium, where
excitations start spreading over a large number of degrees of freedom.
In this case we show some features on the spin dephasing and noise
spectra that manifest the quench --non-stationary-- effects, evidenced
by a bispectrum and a reduced dephasing compared with a stationary
noise. In particular, we show how the quench correlation can be probed
by monitoring this dephasing.

The results presented here thus set a general framework for a large
universal class of non-stationary --out-of-equilibrium-- noise sources
of decoherence, allowing to probe and interpret noise spectral properties
and time-correlations by a quantum probe. Thus they provide tools
and insights for probing and understanding the dynamics of quantum
information of out-of-equilibrium complex quantum systems via a quantum
sensor \citep{Polkovnikov2011,Ma2014,Alvarez2015,Alvarez2015a,Eisert2015,Schweigler2017,Casola2018,Lewis-Swan2019,Wang2019,Davis2021}.
In particular, they can be useful for quantum sensing the dynamics
of single, but complex, large molecules as proteins \citep{Shi2015,Ajoy2015,Lovchinsky2016,Abobeih2019}
and neuronal activity \citep{Hall2012,Hall2013,Zhang2021a} with potential
applications in biology and medicine. At the same time the general
framework sets an universal formula to allow finding optimal control
for protecting against decoherence generated by more realistic environments,
that at atomic scales produce non-stationary --out-of-equilibrium--
noise fluctuations. This tool is very important for implementing quantum
technologies that can span from memory storage to information processing
in quantum devices \citep{Kurizki2015,Tetienne2017,Acin2018,Awschalom2018,Deutsch2020}.
\begin{acknowledgments}
We thank useful discussions with C.D. Fosco and M.J. Sanchez. This
work was supported by CNEA, ANPCyT-FONCyT PICT-2017-3447, PICT-2017-3699,
PICT-2018-04333, PIP-CONICET (11220170100486CO), UNCUYO SIIP Tipo
I 2019-C028, and Instituto Balseiro. M.K. acknowledges support from
Instituto Balseiro (CNEA-UNCUYO). We acknowledge support from CONICET.
\end{acknowledgments}

\appendix

\section{Path integrals for gaussian noise probability distributions\label{sec:Path-integral-for}}

Path integrals can be calculated exactly when the integrand is
determined by Gaussian probability distributions as \citep{zinn}\begin{widetext}
\begin{multline}
\int\mathcal{D}\ensuremath{\mathbf{B}}\exp\left[-\frac{1}{2}(\mathbf{B}|\mathbb{D}|\mathbf{B})+(\mathbf{J}|\mathbf{B})\right]=\int\mathcal{D}\ensuremath{\mathbf{B}}\exp\left[-\frac{1}{2}\int dt\int dt'\mathbf{B}^{\dagger}(t)\mathbb{D}(t,t')\mathbf{B}(t')+\int dt\mathbf{J}^{\dagger}(t)\mathbf{B}(t)\right]\\
=\text{Det}(\nicefrac{\mathbb{D}}{2\pi})^{\nicefrac{1}{2}}\exp\left[\frac{1}{2}\int dt\int dt'\mathbf{J^{\dagger}}(t)\mathbb{G}(t,t^{'})\mathbf{J}(t')\right],\label{eq:gaussian}
\end{multline}
\end{widetext}where $\mathbb{D}$ is any real Hermitian operator,
$\mathbf{J}$ is any function over $\mathfrak{R}^{n}$, and $\mathbb{G}$
is the inverse of the Kernel operator $\mathbb{D}$. The inverse relation
is determined by 
\begin{equation}
\int dt\mathbb{D}(t_{1},t)\,\mathbb{G}(t,t_{2})=\delta(t_{1}-t_{2}),\label{eq:G_definition_}
\end{equation}
where $\delta$ is de Dirac delta distribution and the boundary condition
$\lim_{|t|\to\infty}\mathbb{G}(t,t')=0$ must be satisfied. The path
integral is well defined if and only if $\mathbb{D}$ is a \emph{positive
definite} operator. For these operators, functions such that $\lim_{\left|t\right|\to\infty}\mathbf{B}(t)\neq0$
do not contribute to the integral, as $(\mathbf{B}|\mathbb{D}|\mathbf{B})$
diverges.

Since for any real anti-Hermitian operator $\mathbb{D}^{A}$, $(\mathbf{B}|\mathbb{D}^{A}|\mathbf{B})=0$,
one can only consider integrals where $\mathbb{D}$ is a Hermitian
operator without loss of generality as in Eq. (\ref{eq:gaussian}).
This can be demonstrated by considering that an arbitrary $\mathbb{D}$
can be decomposed as 
\begin{equation}
\mathbb{D}=\mathbb{D}^{H}+\mathbb{D}^{A},
\end{equation}
with $\mathbb{D}^{H}=\frac{\mathbb{D}+\mathbb{D}^{\dagger}}{2}$ and
$\mathbb{D}^{A}=\frac{\mathbb{D}-\mathbb{D}^{\dagger}}{2}$ its Hermitian
and anti-Hermitian parts, respectively. The value of the integral
thus depends only on $\mathbb{D}^{H}$, and therefore we can consider
$\mathbb{D}^{A}=0$. In this article, we have considered $\mathbf{J}=i\mathbf{f}$
and $\mathbf{B}$ the noise fluctuating fields, therefore 
\begin{align}
\langle e^{i\phi[\mathbf{B},\mathbf{f}]}\rangle & =\int\mathcal{D}\ensuremath{\mathbf{B}}\exp\left[-\frac{1}{2}(\mathbf{B}|\mathbb{D}|\mathbf{B})+i(\mathbf{f}|\mathbf{B})\right]\\
 & =\exp\left[-\frac{1}{2}(\mathbf{f}|\mathbb{G}|\mathbf{f})\right].
\end{align}
\medskip{}

\section{Parametrization of the noise eigenmode basis\label{sec:Noise-eigenmode-basis}}

As described in Sec. \ref{subsec:Non-stationary-noise-modes}, $\Omega$
is a parameter that in general its physical meaning depends specifically
on the non-stationary process, and on the parametrization of the functional
behaviour $S(\Omega)$, as it is always possible to reparametrize
the eigenmodes by a different parameter. Similarly, while the frequency
$\omega$ seems a natural choice for stationary environments, different
parameters can be used to describe the noise eigenmodes also in stationary
systems.

Here, we show examples of possible reparametrization of the noise
eigenmodes for stationary noises. Specifically, rather than using
$\Omega=(\omega,m)$ as discussed in Sec. \ref{subsec:Stationary-noise-modes},
we can chose $\Omega'=(\lambda,m)=(\omega^{\nicefrac{1}{3}},m)$.
Therefore the noise spectrum is $S'(\Omega')=S'_{m}(\lambda)=S_{m}(\omega=\lambda^{3})$
and the noise eigenmode $|\Omega')=|\lambda,m)=\sqrt{3}|\lambda|\mathbf{b}_{m}(\omega=\lambda^{3})|\omega=\lambda^{3})$,
where the prefactors are derived from the orthonormal relation $(\lambda,m|\lambda',m')=\delta_{m,m'}\delta(\lambda-\lambda')$,
and that $F'_{m}(\lambda)=(\lambda,m|\mathbf{f})$. Applying these
change of variable $\omega=\lambda^{3}$ to Eq. (\ref{eq:decoherence_overlap_frequency_modes_stationary}),
one obtains 
\[
\langle e^{i\phi[\mathbf{B},\mathbf{f}]}\rangle=\exp\left[-\frac{1}{2}\sum_{m}\int d\lambda\,S'_{m}(\lambda)\left|F'_{m}(\text{\ensuremath{\lambda}})\right|^{2}\right],
\]
which is how Eq. (\ref{eq:Dec_lambda}) is expressed with this new
parametrization. 

Another example is analogous to how the free particle Hamiltonian
eigenstates can be indexed either by their momentum or by their energy
and parity. In this case, the complex exponentials are replaced by
trigonometric functions and the noise eigenmodes are $|E,\pi,m)$.
Here $(t|E,1,m)=\sqrt{\frac{2}{\pi}}\frac{1}{\sqrt[4]{E}}\mathbf{b}_{m}(\omega=\sqrt{E})\cos(\sqrt{E}t)$
and $(t|E,-1,m)=\sqrt{\frac{2}{\pi}}\frac{1}{\sqrt[4]{E}}\mathbf{b}_{m}(\omega=\sqrt{E})\sin(\sqrt{E}t)$,
where the prefactors are again obtained from the normalization condition
$(E,\pi,m|E',\pi',m\text{'})=\delta(E-E')\delta_{\pi,\pi'}\delta_{m,m'}$.

\section{Derivation of the local-in-time differential operator $\mathbb{D}$\label{sec:Simplification-Expression-D}}

The most general action that describes a local-in-time, non-stationary
noise process is of the form 
\begin{equation}
\mathcal{A}[\mathbf{B}]=\int dt\,\sum_{k}\sum_{l\leq k}\mathbf{B}^{\dagger(k)}(t)\mathbb{D}_{k,l}(t)\mathbf{B}^{(l)}(t),
\end{equation}
as in Eq. (\ref{eq:def_operador_diferencial_no_simplificado}) of
the main text. Since the only paths that contribute to the dephasing
integral satisfy $\lim_{|t|\to\infty}\mathbf{B}^{(k)}(t)=0$ for all
$k\leq N$, we integrate by parts each of the action terms and find
that\begin{widetext}
\begin{equation}
\int dt\,\mathbf{B}^{\dagger(k)}(t)\mathbb{D}_{k,l}(t)\mathbf{B}^{(l)}(t)=-\int dt\,\mathbf{B}^{\dagger(k-1)}(t)\left(\mathbb{D}_{k,l}(t)\mathbf{B}^{(l+1)}(t)+\mathbb{\dot{D}}_{k,l}(t)\mathbf{B}^{(l)}(t)\right),
\end{equation}
\end{widetext}where $\dot{\mathbb{D}}{}_{k,l}(t)$ is the derivative
of $\mathbb{D}{}_{k,l}(t)$. Integrating by parts thus transforms
a single term containing the derivatives of order $k$ and $l$ of
$\mathbf{B}$ into two terms, one containing the derivatives of order
$k-1$ and $l+1$ and another containing the ones of order $k-1$
and $l$. Since $l\leq k$, this integration by parts reduces the
difference between the orders of the derivatives when applied to terms
with $k-l>1$. By repeating this procedure successively, one can express
the action only containing terms with $k=l$ and $k=l+1$,
\begin{multline}
\mathcal{A}[\mathbf{B}]=\int dt\,\sum_{k=0}^{N}\mathbf{B}^{\dagger(k)}(t)\mathbb{\tilde{D}}_{k}^{H}(t)\mathbf{B}^{(k)}(t)\\
+\,\sum_{k=1}^{N}\mathbf{B}^{\dagger(k)}(t)\mathbb{\tilde{D}}_{k}^{A}(t)\mathbf{B}^{(k-1)}(t),
\end{multline}
where $\mathbb{\tilde{D}}_{k}^{H}(t)$ and $\mathbb{\tilde{D}}_{k}^{A}(t)$
are real matrices.

Considering that the derivative
\begin{multline}
\frac{d}{dt}\left(\mathbf{B}^{\dagger(k)}(t)\mathbb{\tilde{D}}_{k+1}^{A}(t)\mathbf{B}^{(k)}(t)\right)=\mathbf{B}^{\dagger(k+1)}(t)\mathbb{\tilde{D}}_{k+1}^{A}(t)\mathbf{B}^{(k)}(t)\\
+\mathbf{B}^{\dagger(k)}(t)\mathbb{\tilde{D}}_{k+1}^{A}(t)\mathbf{B}^{(k+1)}(t)+\mathbf{B}^{\dagger(k)}(t)\mathbb{\dot{\tilde{D}}}_{k+1}^{A}(t)\mathbf{B}^{(k)}(t),
\end{multline}
that 
\begin{equation}
\int dt\frac{d}{dt}\left(\mathbf{B}^{\dagger(k)}(t)\mathbb{\tilde{D}}_{k}^{A}(t)\mathbf{B}^{(k)}(t)\right)=0,
\end{equation}
 and that
\begin{equation}
\mathbf{B}^{\dagger(k)}(t)\mathbb{\tilde{D}}_{k+1}^{A}(t)\mathbf{B}^{(k+1)}(t)=\mathbf{B}^{\dagger(k+1)}(t)\mathbb{\tilde{D}}_{k+1}^{A\dagger}(t)\mathbf{B}^{(k)}(t),
\end{equation}
the action can be written as
\begin{multline}
\mathcal{A}[\mathbf{B}]=\int dt\,\sum_{k=0}^{N}\mathbf{B}^{\dagger(k)}(t)\left(\mathbb{\tilde{D}}_{k}^{H}(t)-\frac{1}{2}\mathbb{\dot{\tilde{D}}}_{k+1}^{A}(t)\right)\mathbf{B}^{(k)}(t)\\
+\sum_{k=1}^{N}\frac{1}{2}\mathbf{B}^{\dagger(k)}(t)\left(\mathbb{\tilde{D}}_{k}^{A}(t)-\mathbb{\tilde{D}}_{k}^{A\dagger}(t)\right)\mathbf{B}^{(k-1)}(t),
\end{multline}
where $\mathbb{\dot{\tilde{D}}}_{N+1}^{A}(t)=0$. The terms $\left(\mathbb{D}_{k}^{A}(t)-\mathbb{D}_{k}^{A\dagger}(t)\right)$
are real anti-Hermitian matrices and the value of $\mathbf{B}^{\dagger(k)}(t)\left(\mathbb{\tilde{D}}_{k}^{H}(t)-\frac{1}{2}\mathbb{\dot{\tilde{D}}}_{k+1}^{A}(t)\right)\mathbf{B}^{(k)}(t)$
depends only on the Hermitian part of $\mathbb{\tilde{D}}_{k}^{H}(t)-\frac{1}{2}\mathbb{\dot{\tilde{D}}}_{k}^{A}(t)$.
Therefore, defining  the real Hermitian matrices
\begin{equation}
\mathbb{D}_{k}^{H}=\frac{\mathbb{\tilde{D}}_{k}^{H}(t)-\frac{1}{2}\mathbb{\dot{\tilde{D}}}_{k+1}^{A}(t)+\mathbb{\tilde{D}}_{k}^{H\dagger}(t)-\frac{1}{2}\mathbb{\dot{\tilde{D}}}_{k+1}^{A\dagger}(t)}{2}
\end{equation}
 and the real anti-Hermitian ones
\begin{equation}
\mathbb{D}_{k}^{A}=\frac{\mathbb{D}_{k}^{A}(t)-\mathbb{D}_{k}^{A\dagger}(t)}{2},
\end{equation}
the action can be simplified to

\begin{multline}
\mathcal{A}[\mathbf{B}]=\int dt\,\sum_{k=0}^{N}\mathbf{B}^{\dagger(k)}(t)\mathbb{D}_{k}^{H}(t)\mathbf{B}^{(k)}(t)\\
+\int dt\sum_{k=1}^{N}\mathbf{B}^{\dagger(k)}(t)\mathbb{D}_{k}^{A}(t)\mathbf{B}^{(k-1)}(t).
\end{multline}

Defining then the Hermitian differential operator as 
\begin{multline}
\mathbb{D}=\sum_{k=0}^{N}\overset{\leftarrow}{\partial_{t}^{\,k}}\mathbb{D}_{k}^{H}(t)\,\partial_{t}^{\,k}\\
+\frac{1}{2}\sum_{k=1}^{N}\left[\overset{\leftarrow}{\partial_{t}^{\,k}}\mathbb{D}_{k}^{A}(t)\partial_{t}^{\,k-1}-\overset{\leftarrow}{\partial_{t}^{\,k-1}}\mathbb{D}_{k}^{A}(t)\partial_{t}^{\,k}\right],
\end{multline}
we thus obtain 
\begin{equation}
\mathcal{A}[\mathbf{B}]=\frac{1}{2}(\mathbf{B}|\mathbb{D}|\mathbf{B}),
\end{equation}
as defined in the main text in Eq. (\ref{eq:local-in-time-action}).

The operator $\mathbb{D}$ acts on the field function $\mathbf{B}(t)$
as
\begin{multline}
(t|\mathbb{D}|\boldsymbol{\mathbf{B})}=\sum_{k=0}^{N}(-1)^{k}\partial_{t}^{k}\left[\mathbb{D}_{k}^{H}(t)\partial_{t}^{k}\mathbf{B}(t)\right]\\
+\frac{1}{2}\sum_{k=1}^{N}(-1)^{k}\left[\partial_{t}^{k}\left(\mathbb{D}_{k}^{A}(t)\partial_{t}^{k-1}\mathbf{B}(t)\right)\right]\\
+\frac{1}{2}\sum_{k=1}^{N}(-1)^{k}\left[\partial_{t}^{k-1}\left(\mathbb{D}_{k}^{A}(t)\partial_{t}^{\,k}\mathbf{B}(t)\right)\right],
\end{multline}
where $\overset{\leftarrow}{\partial_{t}}=-\partial_{t}$ over the
space of functions considered, as implementing integration by parts
we find that
\begin{equation}
(f|\overset{\leftarrow}{\partial_{t}}|g)=\int dt\,\dot{f}(t)g(t)=-\int f(t)\dot{g}(t)=-(f|\partial_{t}|g).
\end{equation}
This implies that for a stationary process
\begin{multline}
(\mathbf{B}|\overset{\leftarrow}{\partial_{t}^{k}}\mathbb{D}_{k}^{H}\partial_{t}^{k}|\mathbf{B})=\int dt\mathbf{\mathbf{B}}^{(k)\dagger}(t)\mathbb{D}_{k}^{H}\mathbf{B}^{(k)}(t)\\
=(\mathbf{B}|(-1)^{k}\mathbb{D}_{k}^{H}\partial_{t}^{2k}|\mathbf{B})
\end{multline}
and
\begin{multline}
(\mathbf{B}|\overset{\leftarrow}{\partial_{t}^{k}}\mathbb{D}_{k}^{A}\partial_{t}^{k-1}|\mathbf{B})=\int dt\mathbf{\mathbf{B}}^{(k)\dagger}(t)\mathbb{D}_{k}^{A}\mathbf{B}^{(k-1)}(t)\\
=(\mathbf{B}|(-1)^{k}\mathbb{D}_{k}^{H}\partial_{t}^{2k-1}|\mathbf{B}),
\end{multline}
so that the differential operator is
\begin{equation}
\mathbb{D}=\sum_{k=0}^{N}(-1)^{k}\mathbb{D}_{k}^{H}\,\partial_{t}^{2k}+\sum_{k=1}^{N}(-1)^{k}\mathbb{D}_{k}^{A}\partial_{t}^{\,2k-1}.
\end{equation}

\section{Differentiability of the field paths for local-in-time noises\label{subsec:Differentiability-of-the}}

In this Appendix we demonstrate that the field paths and their first
$N-1$ derivatives must be continuous for local-in-time noise processes.
We thus study the differentiability of the paths allowed by the action.
Considering a particular $\mathbf{B}(t)$, let us assume that at some
time $\overline{t}$ some of the first $N-1$ derivatives of $\mathbf{B}(t)$
are not continuous at $t=\overline{t}$. Then, the $N$ order derivative
\begin{equation}
\mathbf{B}^{(N)}(t)=\mathbf{B}_{0}^{(N)}(t)+\sum_{k=0}^{N-1}\Delta\mathbf{B}^{(k)}\delta^{(N-1-k)}(t-\overline{t}),
\end{equation}
where the coefficients 
\begin{equation}
\Delta\mathbf{B}^{(k)}=\lim_{t\to\overline{t}^{+}}\mathbf{B}^{(k)}(t)-\lim_{t\to\overline{t}^{-}}\mathbf{B}^{(k)}(t)
\end{equation}
for functions for which these limits exist and $\mathbf{B}_{0}^{(N)}(t)$
is a distribution such that $\mathbf{B}_{0}^{(N)}(\overline{t})$
is finite. Then, terms like the following contribute to the action\begin{widetext}
\begin{multline}
\int dt\,\mathbf{B}^{(N)\dagger}(t)\mathbb{D}_{N}^{H}(t)\mathbf{B}^{(N)}(t)=\int dt\,\mathbf{B}_{0}^{(N)\dagger}(t)\mathbb{D}_{N}^{H}(t)\mathbf{B}_{0}^{(N)}(t)+\int dt2\mathbf{B}_{0}^{(N)\dagger}(t)\mathbb{D}_{N}^{H}(t)\sum_{k=0}^{N-1}\Delta\mathbf{B}^{(k)}\delta^{(N-1-k)}(t-\overline{t})+\\
+\int dt\left[\sum_{k=0}^{N-1}\Delta\mathbf{B}^{(k)\dagger}\delta^{(N-1-k)}(t-\overline{t})\right]\mathbb{D}_{N}^{H}(t)\left[\sum_{k=0}^{N-1}\Delta\mathbf{B}^{(k)}\delta^{(N-1-k)}(t-\overline{t})\right].
\end{multline}
\end{widetext} The last term of this integral gives a positively
divergent contribution that depends quadratically on the discontinuity
of $\mathbf{B}(t)$, given by $\Delta\mathbf{B}^{(k)}$, and the first
$N-1$ derivatives of the field at $t=\overline{t}$. The discontinuities
of higher derivatives do not appear in the action. Since the operator
$\mathbb{D}(t)$ is definite positive, the total contribution of these
discontinuous paths to the action must be infinite. Therefore the
probability of discontinuous paths and paths with discontinuities
in their first $N-1$ derivatives, is zero, and do not contribute
to the path integral. Thus we have concluded that the field paths
and its first $N-1$ derivatives must be continuous for local-in-time
processes.

\section{Discretization of local-in-time processes\label{sec:Discretization-local-in-time}}

In this appendix we show an interpretation of the meaning of local-in-time
processes by implementing a time discretization. We replace the continuous
time variable $t$ by the discrete sequence of times $t_{j}=j\Delta t$.
The action is thus
\[
\begin{array}{c}
\mathcal{A}[\boldsymbol{\mathbf{B}}]=\sum_{j}\Delta t\,\sum_{k=0}^{N}\mathbf{B}^{\dagger(k)}(t_{j})\mathbb{D}_{k}^{H}(t_{j})\mathbf{B}^{(k)}(t_{j})\\
+\sum_{j}\Delta t\sum_{k=1}^{N}\mathbf{B}^{\dagger(k)}(t_{j})\mathbb{D}_{k}^{A}(t_{j})\mathbf{B}^{(k-1)}(t_{j}),
\end{array}
\]
where the field derivatives have been replaced using finite differences
$\dot{\mathbf{B}}(t_{j})=\frac{\mathbf{B}(t_{j+1})-\mathbf{B}(t_{j-1})}{2\Delta t}$
with $N$ the highest derivative order that contributes to the action.
Within this discrete representation, the fluctuating field $\mathbf{B}$
can be defined by a vector $\mathbf{B}_{j}=\mathbf{B}(t_{j})$ and
the kernel operator $\mathbb{D}$ can be represented by the matrix
$\mathbb{D}_{jj'}=\mathbb{D}(t_{j},t_{j'})$. Therefore, the $k$-th
derivative of the field $\left[\mathbf{B}^{(k)}\right]_{j}=\sum_{j'=j-k}^{j+k}\frac{c_{k}}{\Delta t^{k}}\mathbf{B}_{j'}$
is a linear combination of the values that $\mathbf{B}$ takes at
the $2k+1$ times closest to $j$, where the coefficients $c_{k}$
are independent of $\Delta t$. Thus, a kernel operator corresponding
to a local-in-time process has coefficients that satisfy $\mathbb{D}_{jj'}=0$
for $|j-j'|>N$. Within this discrete picture, local-in-time processes
are those whose kernel operator is $N$-diagonal, i.e. its matrix
representation has non zero coefficients only in the first $N$ central
diagonals. In this representation only if $N=0$, the kernel operator
is diagonal in the time basis and therefore $\mathbf{B}(t)$ and $\mathbf{B}(t')$
are uncorrelated for $t\neq t'$. However for $N\geq1$ the field
at the $N$-th closest times are correlated. Still, local-in-time
processes are such that long-time correlations do not exist. 

\medskip{}

\section{Non-Markovian proof for $N>1$\label{subsec:Non-Markovian-proof-for}\label{sec:Non-markovian-noise-demonstratio}}

In this Appendix we analyze the conditions for a non-stationary
noise to be non-Markovian. As we stated in the main text, noise processes
that are not local-in-time cannot be Markovian since the action explicitly
relates the value of the fluctuating field $\mathbf{B}$ at different
times. A noise with $N=0$ is the so-called white noise, and therefore
it is Markovian. Therefore, we only consider in the following demonstration,
local-in-time noises with $N\geq1$. In Appendix \ref{sec:Markovian-full-stochastic}
we prove that the generalized noise process given by $\{\mathbf{B}^{(k)}(t)\}$
is Markovian. As a particular case, if $N=1$, then $\{\mathbf{B}^{(k)}(t)\}\equiv\{\mathbf{B}^{(k)}(t)\,:\,k=0\}=\{\mathbf{B}(t)\}$,
therefore showing that processes with $N=1$ are Markovian. In the
Appendix \ref{subsec:Differentiability-of-the}, we prove that the
field paths that contribute to the path integral for local-in-time
noises are not only continuous, but are $N-1$ times continuously
differentiable. This property allows us to demonstrate that noises
with $N>1$ are not Markovian.

To do this, we consider the values of the field at two infinitesimally
close times $t_{0}$ and $t_{1}$ to be $\mathbf{B}_{0}$ and $\mathbf{B}_{1}$,
respectively. According to this assumption, the field derivative $\dot{\mathbf{B}}(t_{1})=\frac{\mathbf{\mathbf{B}_{1}-\mathbf{B}_{0}}}{t_{1}-t_{0}}$
must be continuous. If we now consider the field value at a time $t_{f}$,
infinitesimally close and after $t_{1}$, we know again that $\dot{\mathbf{B}}$
will change continuously. Therefore, the mean value of the probability
distribution $P(\mathbf{B}_{f},t_{f}|\mathbf{B}_{1},t_{1};\mathbf{B}_{0},t_{0})$
is $\mathbf{B}_{1}+\dot{\mathbf{B}}(t_{1})\left(t_{f}-t_{1}\right)=\mathbf{\mathbf{B}}_{1}\left(1+\frac{t_{f}-t_{1}}{t_{1}-t_{0}}\right)-\mathbf{B}_{0}\frac{t_{f}-t_{1}}{t_{1}-t_{0}}$.
This means that the state of the field at $t_{f}$ depends on the
state $\mathbf{B}_{0}$ and $\mathbf{B}_{1}$ at times $t_{0}$ and
$t_{1}$ respectively. Therefore, the probability distribution of
the field depends on the state of the system at least at two previous
times, thus 
\begin{equation}
P(\mathbf{B}_{f},t_{f}|\mathbf{B}_{1},t_{1};\mathbf{B}_{0},t_{0})\neq P(\mathbf{B}_{f},t_{f}|\mathbf{B}_{1},t_{1}).
\end{equation}
A Markovian process must satisfy Eq. (\ref{eq:prop_Markov}), therefore
in this case we have that 
\begin{equation}
\begin{array}{c}
P(\mathbf{B}_{f},t_{f}|\mathbf{B}_{0},t_{0})=\\
\int d\mathbf{B}_{1}P(\mathbf{B}_{f},t_{f}|\mathbf{B}_{1},t_{1};\mathbf{B}_{0},t_{0})P(\mathbf{B}_{1},t_{1}|\mathbf{B}_{0},t_{0})\\
\neq\int d\mathbf{B}_{1}P(\mathbf{B}_{f},t_{f}|\mathbf{B}_{1},t_{1})P(\mathbf{B}_{1},t_{1}|\mathbf{B}_{0},t_{0}),
\end{array}
\end{equation}
and thus the process is not Markovian. This example evidences the
key considerations to show why a local-in-time noise process is not
Markovian if the differential operator $\mathbb{D}(t)$ contains derivatives
higher than the order $N=1$. In this case of $N>1$, the state at
a given time depends on the information about previous states that
is encoded on the continuous derivatives of the field.

\section{Markovian generalized stochastic process \label{sec:Markovian-full-stochastic}}

In this Appendix, we find a formula for the propagator of the generalized
noise process
\begin{equation}
\begin{array}{cc}
\{\mathbf{B}^{(k)}(t)\} & \equiv\{\mathbf{B}^{(k)}(t)\,:\,k=0,\dots N-1\}\\
 & =\{\mathbf{B}(t),\dot{\mathbf{B}}(t),\dots,\mathbf{B}^{(N-1)}(t)\},
\end{array}
\end{equation}
introduced in Sec. \textcolor{red}{\ref{sec:local-in-time}} of the
main text, in terms of path integrals. We show that it can be separated
into three independent path integrals with boundary conditions. We
use this integral decomposition to prove that the process that describes
$\{\mathbf{B}^{(k)}(t)\}$ is Markovian.

We consider how the probability distribution of the generalized field
fluctuations $\{\mathbf{B}^{(k)}(t)\}$ evolves with time. To demonstrate
the Markovian condition of Eq. (\ref{eq:prop_Markov}) of the main
text, we need to calculate the probability of the state $\{\mathbf{B}_{f}^{(k)}\}$
at time $t_{f}$ given that at time $t_{0}$ the field and its derivatives
are $\{\mathbf{B}_{0}^{(k)}\}$. This conditional probability is given
by
\begin{equation}
\begin{aligned}P(\{\mathbf{B}_{f}^{(k)}\},t_{f}| & \{\mathbf{B}_{0}^{(k)}\}t_{0})=\\
\frac{1}{P(\{\mathbf{B}_{0}^{(k)}\},t_{0})}\int_{\left\{ \mathbf{B}|\{\mathbf{B}^{(k)}(t_{0,f})\}=\{\mathbf{B}_{0,f}^{(k)}\}\right\} }\!\!\!\!\!\!\!\! & \mathcal{D}\mathbf{B}\exp\left[-\frac{1}{2}(\mathbf{B}|\mathbb{D}|\mathbf{B})\right],
\end{aligned}
\label{eq:P(=00005BB_f=00005D,t_f|=00005BB_0=00005D,t_0)-Int-1}
\end{equation}
where the path integral runs over all the paths $\mathbf{B}(t)$ such
that $\{\mathbf{B}^{(k)}(t_{0})\}=\{\mathbf{B}_{0}^{(k)}\}$ and $\{\mathbf{B}^{(k)}(t_{f})\}=\{\mathbf{B}_{f}^{(k)}\}$,
denoted by $\left\{ \mathbf{B}|\{\mathbf{B}^{(k)}(t_{0,f})\}=\{\mathbf{B}_{0,f}^{(k)}\}\right\} $
in the integral.

We first determine the probability of the state 
\begin{equation}
\{\mathbf{B}_{0}^{(k)}\}\equiv\{\mathbf{B}_{0},\dots,\mathbf{B}_{0}^{(N-1)}\}
\end{equation}
 at time $t_{0}$, which is given by 
\begin{equation}
P(\{\mathbf{B}_{0}^{(k)}\},t_{0})\!=\!\!\int_{\{\mathbf{B}|\{\mathbf{B}^{(k)}(t_{0})\}=\{\mathbf{B}_{0}^{(k)}\}\}}\mathcal{\!\!\!\!\!\!\!\!\!\!\!\!\!\!D}\mathbf{B}\exp\left[-\frac{1}{2}(\mathbf{B}|\mathbb{D}|\mathbf{B})\right],
\end{equation}
where the integral runs over all paths that pass through $\{\mathbf{B}_{0}^{(k)}\}$
at time $t=t_{0}$. The continuity conditions of $\{\mathbf{B}^{(k)}\}$
demonstrated in Appendix \ref{subsec:Differentiability-of-the} implies
that integrating over all paths such that $\{\mathbf{B}^{(k)}(t_{0})\}=\{\mathbf{B}_{0}^{(k)}\}$
is the same than integrating over all paths that end at $t_{0}$ with
$\{\mathbf{B}^{(k)}(t_{0})\}=\{\mathbf{B}_{0}^{(k)}\}$, and over
all paths that start at $t_{0}$ with $\{\mathbf{B}^{(k)}(t_{0})\}=\{\mathbf{B}_{0}^{(k)}\}$,
and then multiplying these two integrals. Therefore, the probability
\begin{equation}
P(\{\mathbf{B}_{0}^{(k)}\},t_{0})=\mathcal{I}_{\left(0,-\infty\right)}^{\left(\left\{ \mathbf{B}_{0}^{(k)}\right\} ,t_{0}\right)}\times\mathcal{I}_{\left(\left\{ \mathbf{B}_{0}^{(k)}\right\} ,t_{0}\right)}^{\left(0,+\infty\right)},\label{eq:P(=00007BB=00007D,t)-separada-en-2}
\end{equation}
where we introduced the integrals with boundary conditions for the
paths 
\begin{multline}
\mathcal{I}_{\left(\left\{ \mathbf{B}_{0}^{(k)}\right\} ,t_{0}\right)}^{\left(0,+\infty\right)}\equiv\\
\int_{\left(\left\{ \mathbf{B}_{0}^{(k)}\right\} ,t_{0}\right)}^{\left(0,+\infty\right)}\!\!\!\!\!\!\!\!\mathcal{D}\mathbf{B}\exp\left[-\frac{1}{2}\int_{t_{0}}^{+\infty}dt\mathbf{B}^{\dagger}(t)\mathbb{D}(t)\mathbf{B}(t)\right],\label{eq:def_I_t0^inf}
\end{multline}
where the integral runs over all paths that start at time $t_{0}$
with value $\{\mathbf{B}_{0}^{(k)}\}$ and go to $t\to+\infty$ with
$\lim_{t\to\infty}\left\{ \mathbf{B}^{(k)}(t)\right\} =0$, and 
\begin{multline}
\mathcal{I}_{\left(0,-\infty\right)}^{\left(\left\{ \mathbf{B}_{0}^{(k)}\right\} ,t_{0}\right)}\!\equiv\\
\int_{\left(0,-\infty\right)}^{(\left\{ \mathbf{B}_{0}^{(k)}\right\} ,t_{0})}\mathcal{\!\!\!\!\!\!\!\!D}\mathbf{B}\exp\left[-\frac{1}{2}\int_{-\infty}^{t_{0}}dt\mathbf{B}^{\dagger}(t)\mathbb{D}(t)\mathbf{B}(t)\right],\label{eq:def_I_-inf^tf}
\end{multline}
where the integral runs over all the paths that come from $t\to-\infty$
with $\lim_{t\to-\infty}\left\{ \mathbf{B}^{(k)}(t)\right\} =0$,
and end at time $t_{0}$ with value $\{\mathbf{B}_{0}^{(k)}\}$.

We then calculate the integral on the numerator of Eq. (\ref{eq:P(=00005BB_f=00005D,t_f|=00005BB_0=00005D,t_0)-Int-1}),
which gives 
\begin{equation}
\begin{array}{c}
\int_{\{\mathbf{B}|\{\mathbf{B}^{(k)}(t_{0,f})\}=\{\mathbf{B}_{0,f}^{(k)}\}\}}\mathcal{D}\mathbf{B}\exp\left[-\frac{1}{2}(\mathbf{B}|\mathbb{D}|\mathbf{B})\right]=\\
\mathcal{I}_{\left(0,-\infty\right)}^{\left(\left\{ \mathbf{B}_{0}^{(k)}\right\} ,t_{0}\right)}\times\mathcal{I}_{\left(\left\{ \mathbf{B}_{0}^{(k)}\right\} ,t_{0}\right)}^{\left(\left\{ \mathbf{B}_{f}^{(k)}\right\} ,t_{f}\right)}\times\mathcal{I}_{\left(\left\{ \mathbf{B}_{f}^{(k)}\right\} ,t_{f}\right)}^{\left(0,+\infty\right)},
\end{array}\label{eq:P(=00007BB_f=00007D,t_f;=00007BB_0=00007D,t_0)-separada-en-3}
\end{equation}
where we have introduced the integral with boundary conditions 
\begin{multline}
\mathcal{I}_{\left(\left\{ \mathbf{B}_{0}^{(k)}\right\} ,t_{0}\right)}^{\left(\left\{ \mathbf{B}_{f}^{(k)}\right\} ,t_{f}\right)}\!\equiv\\
\int_{\left(\left\{ \mathbf{B}_{0}^{(k)}\right\} ,t_{0}\right)}^{\left(\left\{ \mathbf{B}_{0}^{(k)}\right\} ,t_{0}\right)}\mathcal{\!\!\!\!\!\!\!\!D}\mathbf{B}\exp\left[-\frac{1}{2}\int_{t_{0}}^{t_{f}}dt\mathbf{B}^{\dagger}(t)\mathbb{D}(t)\mathbf{B}(t)\right],\label{eq:def_I_t0^tf}
\end{multline}
where the integral runs over all the paths that start at time $t_{0}$
with value $\{\mathbf{B}_{0}^{(k)}\}$ and end at time $t_{f}$ with
value $\{\mathbf{B}_{f}^{(k)}\}$. Again, we have considered the continuity
conditions on the field paths $\{\mathbf{B}^{(k)}\}$ demonstrated
in Appendix \ref{subsec:Differentiability-of-the}. The integrals
$\mathcal{I}_{\left(\left\{ \mathbf{B}_{f}^{(k)}\right\} ,t_{f}\right)}^{\left(0,+\infty\right)}$
and $\mathcal{I}_{\left(0,-\infty\right)}^{\left(\left\{ \mathbf{B}_{0}^{(k)}\right\} ,t_{0}\right)}$
are defined in Eqs. (\ref{eq:def_I_t0^inf}) and (\ref{eq:def_I_-inf^tf})
respectively.

Now using Eqs. (\ref{eq:P(=00007BB=00007D,t)-separada-en-2}) and
(\ref{eq:P(=00007BB_f=00007D,t_f;=00007BB_0=00007D,t_0)-separada-en-3}),
we obtain the conditional probability of the noise process

\begin{equation}
P(\{\mathbf{B}_{f}^{(k)}\},t_{f}|\{\mathbf{B}_{0}^{(k)}\}|t_{0})=\frac{\mathcal{I}_{\left(\left\{ \mathbf{B}_{0}^{(k)}\right\} ,t_{0}\right)}^{\left(\left\{ \mathbf{B}_{f}^{(k)}\right\} ,t_{f}\right)}\times\mathcal{I}_{\left(\left\{ \mathbf{B}_{f}^{(k)}\right\} ,t_{f}\right)}^{\left(0,+\infty\right)}}{\mathcal{I}_{\left(\left\{ \mathbf{B}_{0}^{(k)}\right\} ,t_{0}\right)}^{\left(0,+\infty\right)}}.\label{eq:P(=00005BB_f=00005D,t_f|=00005BB_0=00005D,t_0)-Apx-1}
\end{equation}
In order to prove that this process is Markovian, we introduce an
intermediate time $t_{1}$ and calculate the integral

\begin{multline}
\!\int\!\!d\{\mathbf{B}_{1}^{(k)}\!\}P(\{\mathbf{B}_{f}^{(k)}\!\},t_{f}|\{\mathbf{B}_{1}^{(k)}\!\},t_{1})P(\{\mathbf{B}_{1}^{(k)}\!\},t_{1}|\{\mathbf{B}_{0}^{(k)}\!\},t_{0})\\
\!=\!\!\int\!\!d\{\mathbf{B}_{1}^{(k)}\!\}\times\\
\times\frac{\mathcal{I}_{\left(\left\{ \mathbf{B}_{0}^{(k)}\!\right\} ,t_{0}\right)}^{\left(\left\{ \mathbf{B}_{1}^{(k)}\!\right\} ,t_{1}\right)}\times\mathcal{I}_{\left(\left\{ \mathbf{B}_{1}^{(k)}\!\right\} ,t_{1}\right)}^{\left(0,+\infty\right)}}{\mathcal{I}_{\left(\left\{ \mathbf{B}_{0}^{(k)}\!\right\} ,t_{0}\right)}^{\left(0,+\infty\right)}}\frac{\mathcal{I}_{\left(\left\{ \mathbf{B}_{1}^{(k)}\!\right\} ,t_{1}\right)}^{\left(\left\{ \mathbf{B}_{f}^{(k)}\!\right\} ,t_{f}\right)}\times\mathcal{I}_{\left(\left\{ \mathbf{B}_{f}^{(k)}\!\right\} ,t_{f}\right)}^{\left(0,+\infty\right)}}{\mathcal{I}_{\left(\left\{ \mathbf{B}_{1}^{(k)}\!\right\} ,t_{1}\right)}^{\left(0,+\infty\right)}}\\
=\frac{\mathcal{I}_{\left(\left\{ \mathbf{B}_{f}^{(k)}\right\} ,t_{f}\right)}^{\left(0,+\infty\right)}}{\mathcal{I}_{\left(\left\{ \mathbf{B}_{0}^{(k)}\right\} ,t_{0}\right)}^{\left(0,+\infty\right)}}\int d\{\mathbf{B}_{1}^{(k)}\}\mathcal{I}_{\left(\left\{ \mathbf{B}_{0}^{(k)}\right\} ,t_{0}\right)}^{\left(\left\{ \mathbf{B}_{1}^{(k)}\right\} ,t_{1}\right)}\times\mathcal{I}_{\left(\left\{ \mathbf{B}_{1}^{(k)}\right\} ,t_{1}\right)}^{\left(\left\{ \mathbf{B}_{f}^{(k)}\right\} ,t_{f}\right)},
\end{multline}
that appears in the Markovian condition of Eq. (\ref{eq:prop_Markov}),
where $d\{\mathbf{B}_{1}^{(k)}\}\equiv d\mathbf{B}_{1}d\dot{\mathbf{B}_{1}}\dots d\mathbf{B}^{(N-1)}$.

The continuity conditions for $\{\mathbf{B}^{(k)}\}$ at $t_{1}$
imply that $\mathbf{B}^{(k)}(t_{1}^{-})=\mathbf{B}^{(k)}(t_{1}^{+})$
for $0\leq k\leq N-1$. Then this continuity conditions imply that
the integral
\begin{equation}
\int d\{\mathbf{B}_{1}^{(k)}\}\mathcal{I}_{\left(\left\{ \mathbf{B}_{0}^{(k)}\right\} ,t_{0}\right)}^{\left(\left\{ \mathbf{B}_{1}^{(k)}\right\} ,t_{1}\right)}\times\mathcal{I}_{\left(\left\{ \mathbf{B}_{1}^{(k)}\right\} ,t_{1}\right)}^{\left(\left\{ \mathbf{B}_{f}^{(k)}\right\} ,t_{f}\right)}=\mathcal{I}_{\left(\left\{ \mathbf{B}_{0}^{(k)}\right\} ,t_{0}\right)}^{\left(\left\{ \mathbf{B}_{f}^{(k)}\right\} ,t_{f}\right)},
\end{equation}
and thus

\begin{equation}
\begin{array}{c}
\int d\{\mathbf{B}_{1}^{(k)}\!\}P(\{\mathbf{B}_{f}^{(k)}\!\},t_{f}|\{\mathbf{B}_{1}^{(k)}\!\},t_{1})P(\{\mathbf{B}_{1}^{(k)}\!\},t_{1}|\{\mathbf{B}_{0}^{(k)}\!\},t_{0})\\
=\frac{\mathcal{I}_{\left(\left\{ \mathbf{B}_{f}^{(k)}\right\} ,t_{f}\right)}^{\left(0,+\infty\right)}}{\mathcal{I}_{\left(\left\{ \mathbf{B}_{0}^{(k)}\right\} ,t_{0}\right)}^{\left(0,+\infty\right)}}\mathcal{I}_{\left(\left\{ \mathbf{B}_{0}^{(k)}\right\} ,t_{0}\right)}^{\left(\left\{ \mathbf{B}_{f}^{(k)}\right\} ,t_{f}\right)}\\
=P(\{\mathbf{B}_{f}^{(k)}\},t_{f}|\{\mathbf{B}_{0}^{(k)}\},t_{0}),
\end{array}
\end{equation}
demonstrating that the generalized noise process is Markovian.

\section{Markovian propagator for the generalized stochastic process\label{sec:Markovian-Propagator}}

Since local-in-time noises can be described by the generalized Markovian
process of $\{\mathbf{B}^{(k)}(t)\}$, its \emph{propagator} $P(\{\mathbf{B}_{f}^{(k)}\},t_{f}|\{\mathbf{B}_{0}^{(k)}\},t_{0})$
contains all the \emph{information} about the stochastic process that
describes the field paths $\mathbf{B}$. The path integral framework
allows to calculate the \emph{propagator} of $\{\mathbf{B}^{(k)}\}$
without actually requiring to perform \emph{path integrals}. Instead
it can be obtained by solving an \emph{ordinary linear differential
equation }with three different types of \emph{boundary conditions.
}The propagator can be expressed as 
\begin{equation}
P(\{\mathbf{B}_{f}^{(k)}\},t_{f}|\{\mathbf{B}_{0}^{(k)}\}|t_{0})=\frac{\mathcal{I}_{\left(\left\{ \mathbf{B}_{0}^{(k)}\right\} ,t_{0}\right)}^{\left(\left\{ \mathbf{B}_{f}^{(k)}\right\} ,t_{f}\right)}\times\mathcal{I}_{\left(\left\{ \mathbf{B}_{f}^{(k)}\right\} ,t_{f}\right)}^{\left(0,+\infty\right)}}{\mathcal{I}_{\left(\left\{ \mathbf{B}_{0}^{(k)}\right\} ,t_{0}\right)}^{\left(0,+\infty\right)}},
\end{equation}
according to Eq. (\ref{eq:P(=00005BB_f=00005D,t_f|=00005BB_0=00005D,t_0)-Apx-1}).

In order to obtain the explicit formula for the propagator for local-in-time
non-stationary noises, we only need to calculate the integrals $\mathcal{I}_{\left(\left\{ \mathbf{B}_{0}^{(k)}\right\} ,t_{0}\right)}^{\left(\left\{ \mathbf{B}_{f}^{(k)}\right\} ,t_{f}\right)}$,
$\mathcal{I}_{\left(\left\{ \mathbf{B}_{0}^{(k)}\right\} ,t_{0}\right)}^{\left(0,+\infty\right)}$,
and $\mathcal{I}_{\left(\left\{ \mathbf{B}_{f}^{(k)}\right\} ,t_{f}\right)}^{\left(0,+\infty\right)}$.
To do this, we introduce the classical field $\mathbf{B}_{cl}$, which
is defined as the path that minimizes the action with the fixed boundary
conditions. Using variational analysis, one can find that $\mathbf{B}_{cl}$
is the solution of the following differential equation 
\begin{align}
\sum_{k=0}^{N}(-1)^{k}\partial_{t}^{k} & \left[\mathbb{D}_{k}^{H}(t)\partial_{t}^{k}\mathbf{B}_{cl}(t)\right]\label{eq:def_B_cl-appendix}\\
+\frac{1}{2}\sum_{k=1}^{N}(-1)^{k} & \left[\partial_{t}^{k}\left(\mathbb{D}_{k}^{A}(t)\partial_{t}^{k-1}\mathbf{B}_{cl}(t)\right)\right]\nonumber \\
+\frac{1}{2}\sum_{k=1}^{N}(-1)^{k} & \left[\partial_{t}^{k-1}\left(\mathbb{D}_{k}^{A}(t)\partial_{t}^{\,k}\mathbf{B}_{cl}(t)\right)\right]=0,\nonumber 
\end{align}
 with the corresponding boundary conditions for each of the integrals.
They are 
\begin{equation}
\{\mathbf{B}_{cl}^{(k)}(t_{0})\}=\{\mathbf{B}_{0}^{(k)}\},\label{eq:=00005BB_cl(0)=00005D-appendix}
\end{equation}
\begin{equation}
\{\mathbf{B}_{cl}^{(k)}(t_{f})\}=\{\mathbf{B}_{f}^{(k)}\}\label{eq:=00005BB_cl(f)=00005D-appendix}
\end{equation}
 for $\mathcal{I}_{\left(\left\{ \mathbf{B}_{0}^{(k)}\right\} ,t_{0}\right)}^{\left(\left\{ \mathbf{B}_{f}^{(k)}\right\} ,t_{f}\right)}$,
\begin{equation}
\{\mathbf{B}_{cl}^{(k)}(t_{0})\}=\{\mathbf{B}_{0}^{(k)}\},\label{eq:=00005BB_cl(0)=00005D-appendix-1}
\end{equation}
\begin{equation}
\lim_{t\to+\infty}\{\mathbf{B}_{cl}^{(k)}(t)\}=0\label{eq:=00005BB_cl(f)=00005D-appendix-1}
\end{equation}
 for $\mathcal{I}_{\left(\left\{ \mathbf{B}_{0}^{(k)}\right\} ,t_{0}\right)}^{\left(0,+\infty\right)}$,
and
\begin{equation}
\{\mathbf{B}_{cl}^{(k)}(t_{f})\}=\{\mathbf{B}_{f}^{(k)}\},\label{eq:=00005BB_cl(0)=00005D-appendix-2}
\end{equation}
\begin{equation}
\lim_{t\to+\infty}\{\mathbf{B}_{cl}^{(k)}(t)\}=0\label{eq:=00005BB_cl(f)=00005D-appendix-2}
\end{equation}
 for $\mathcal{I}_{\left(\left\{ \mathbf{B}_{f}^{(k)}\right\} ,t_{f}\right)}^{\left(0,+\infty\right)}$.
Notice that Eq. (\ref{eq:def_B_cl-appendix}) is a linear differential
equation of degree $2N$, and the boundary constraints provide the
necessary conditions for the solution to be unique. For the case of
$\mathcal{I}_{\left(\left\{ \mathbf{B}_{0}^{(k)}\right\} ,t_{0}\right)}^{\left(\left\{ \mathbf{B}_{f}^{(k)}\right\} ,t_{f}\right)}$,
the classical field $\mathbf{B}_{cl}(t)$ is the path that minimizes
the action $\mathcal{A}[\mathbf{B}]$ of all paths between $t_{0}$
and $t_{f}$ with fixed endpoints at the bounds of the integral. The
classical field encodes the dependency of the integral on the boundary
conditions $\left(\{\mathbf{B}_{0}^{(k)}\},t_{0}\right)$, and $\left(\{\mathbf{B}_{f}^{(k)}\},t_{f}\right)$.
For $\mathcal{I}_{\left(\left\{ \mathbf{B}_{f}^{(k)}\right\} ,t_{f}\right)}^{\left(0,+\infty\right)}$
and $\mathcal{I}_{\left(\left\{ \mathbf{B}_{0}^{(k)}\right\} ,t_{0}\right)}^{\left(0,+\infty\right)}$,
the classical field is the path that minimizes the action over the
corresponding time intervals, and encodes the dependency on the corresponding
boundary conditions.

By performing the change of variables $\mathbf{B}(t)=\Delta\mathbf{B}(t)+\mathbf{B}_{cl}(t)$,
we can write the integral
\begin{equation}
\mathcal{I}_{\left(\left\{ \mathbf{B}_{0}^{(k)}\right\} ,t_{0}\right)}^{\left(\left\{ \mathbf{B}_{f}^{(k)}\right\} ,t_{f}\right)}=\int_{(\{\Delta\mathbf{B}^{(k)}\}=0,t_{0})}^{(\{\Delta\mathbf{B}^{(k)}\}=0,t_{f})}\mathcal{D}\Delta\mathbf{B}\,e^{-\mathcal{A}[\Delta\mathbf{B}+\mathbf{B}_{cl}]},\label{eq:donde_usamos_shift_invariance}
\end{equation}
where the integral runs over all paths that start at $t_{0}$ and
end at $t_{f}$ with value $\{\Delta\mathbf{B}^{(k)}\}=0$ and we
have considered that the path integral is \emph{invariant under shifts}.
 Notice that the limits of the path integral do not depend on the
value of the fields at $t_{0}$ and $t_{f}$. Similar expression are
obtained for the other two integrals.

The action can thus be written as
\begin{equation}
\begin{array}{c}
\mathcal{A}[\mathbf{B}_{cl}+\Delta\mathbf{B}]=\mathcal{A}[\mathbf{B}_{cl}]+\mathcal{A}[\Delta\mathbf{B}]+\\
+2\int_{t_{0}}^{t_{f}}\Delta\mathbf{B}(t)\mathbb{D}(t)\mathbf{B}_{cl}(t).
\end{array}\label{eq:Acci=0000F3n_separada_en_3}
\end{equation}
The Eq. (\ref{eq:def_B_cl-appendix}) holds if and only if $\int_{t_{0}}^{t_{f}}\Delta\mathbf{B}(t)\mathbb{D}(t)\mathbf{B}_{cl}(t)=0$
for any $\Delta\mathbf{B}(t)$ with $\{\Delta\mathbf{B}^{(k)}(t_{f})\}=\{\Delta\mathbf{B}^{(k)}(t_{0})\}=0$.
Therefore the last term in Eq. (\ref{eq:Acci=0000F3n_separada_en_3})
vanishes and we obtain 
\begin{equation}
\mathcal{A}[\mathbf{B}_{cl}+\Delta\mathbf{B}]=\mathcal{A}[\mathbf{B}_{cl}]+\mathcal{A}[\Delta\mathbf{B}]
\end{equation}
 and the integral becomes
\begin{equation}
\mathcal{I}_{\left(\left\{ \mathbf{B}_{0}^{(k)}\right\} ,t_{0}\right)}^{\left(\left\{ \mathbf{B}_{f}^{(k)}\right\} ,t_{f}\right)}=\int_{(\{\Delta\mathbf{B}^{(k)}\}=0,t_{0})}^{(\{\Delta\mathbf{B}^{(k)}\}=0,t_{f})}\mathcal{D}\Delta\mathbf{B}\,e^{-(\mathcal{A}[\Delta\mathbf{B}]+\mathcal{A}[\mathbf{B}_{cl}])}\,.
\end{equation}
We were thus able to separate the original path integral into two
parts: a path integral with no dependence on the initial and final
fields 
\begin{equation}
M_{t_{0}}^{t_{f}}=\int_{(\{\Delta\mathbf{B}^{(k)}\}=0,t_{0})}^{(\{\Delta\mathbf{B}^{(k)}\}=0,t_{f})}\mathcal{D}\Delta\mathbf{B}\,e^{-\mathcal{A}[\Delta\mathbf{B}]},\label{eq:factor_normalizaci=0000F3n_I}
\end{equation}
and a term without path integrals that does depend on the initial
and final fields 
\begin{equation}
e^{-\mathcal{A}[\mathbf{B}_{cl}]}.
\end{equation}
We have that $M_{t_{0}}^{t_{f}}$ is just a constant that depends
on the initial and final times but not on the initial and final fields.
Since $M_{t_{0}}^{t_{f}}$ does not depends on the values of $\{\mathbf{B}_{0}^{(k)}\}$
and $\{\mathbf{B}_{f}^{(k)}\}$, its effect on the propagator is that
of a renormalization constant that can be calculated by demanding
the normalization of the conditional probability $\int d\{\mathbf{B}_{f}\}P(\{\mathbf{B}_{f}\},t_{f}|\{\mathbf{B}_{0}\},t_{0})=1$.
Then, if we define the classical action
\begin{equation}
\mathcal{A}_{cl}\left(\left\{ \mathbf{B}_{f}^{(k)}\right\} ,t_{f};\left\{ \mathbf{B}_{0}^{(k)}\right\} ,t_{0}\right)\equiv\mathcal{A}[\mathbf{B}_{cl}],\label{eq:Acci=0000F3n_cl=0000E1sica-1}
\end{equation}
the integral

\begin{equation}
\mathcal{I}_{\left(\left\{ \mathbf{B}_{0}^{(k)}\right\} ,t_{0}\right)}^{\left(\left\{ \mathbf{B}_{f}^{(k)}\right\} ,t_{f}\right)}=M_{t_{0}}^{t_{f}}\,e^{-\mathcal{A}_{cl}(\{\mathbf{B}_{f}\},t_{f};\{\mathbf{B}_{0}\},t_{0})}
\end{equation}
is just a normal integral based on the solution of the classical field
for the ordinary differential equation. Analogous results hold for
$\mathcal{I}_{\left(\left\{ \mathbf{B}_{0}^{(k)}\right\} ,t_{0}\right)}^{\left(0,+\infty\right)}$
and $\mathcal{I}_{\left(\left\{ \mathbf{B}_{f}^{(k)}\right\} ,t_{f}\right)}^{\left(0,+\infty\right)}$,
where one must solve the same differential equation, but changing
the boundary conditions.

Therefore we have shown here that the problem of calculating the Markovian
propagator reduces to solve an ordinary differential equation with
three different boundary conditions up to a normalization constant.
Its solution is thus given by

\begin{multline*}
P(\{\mathbf{B}_{f}^{(k)}\},t_{f}|\{\mathbf{B}_{0}^{(k)}\}|t_{0})=\frac{M_{t_{0}}^{t_{f}}\times M_{t_{f}}^{\infty}}{M_{t_{0}}^{\infty}}\times\\
e^{-\left[\mathcal{A}_{cl}(\{\mathbf{B}_{f}\},t_{f};\{\mathbf{B}_{0}\},t_{0})+\mathcal{A}_{cl}(0,+\infty;\{\mathbf{B}_{f}\},t_{f})-\mathcal{A}_{cl}(0,+\infty;\{\mathbf{B}_{0}\},t_{0})\right]},
\end{multline*}
where $\frac{M_{t_{0}}^{t_{f}}\times M_{t_{f}}^{\infty}}{M_{t_{0}}^{\infty}}$
is a normalization constant.

\bibliographystyle{apsrev4-1}
\bibliography{biblio-2}

\end{document}